\documentclass[prx,twocolumn,10pt,aps,longbibliography]{revtex4-2}
\usepackage{amsmath,slashed,amsfonts,graphicx,verbatim}
\usepackage{xcolor}
\usepackage{notes2bib}

\usepackage[utf8]{inputenc}
\usepackage{bm,physics,amssymb,xcolor,graphicx}
\usepackage{setspace}
\usepackage{accents}
\usepackage[top=1in,bottom=1in,left=1in,right=1in]{geometry}
\usepackage{xcolor}
\usepackage{comment}
\usepackage{enumitem}
\usepackage{graphicx} 
\usepackage{verbatim}
\usepackage[mathscr]{euscript}

\usepackage{soul}
\usepackage{caption}
\usepackage{subcaption}

\begin{document}

\title{Dynamically Robust Counterdiabatic Topological Pumping}
\author{Joshua Chiel}
\affiliation{ Department of Physics, University of Maryland, College Park, Maryland 20742, US}
\author{Christopher Jarzynski}
\affiliation{Department of Physics, University of Maryland, College Park, Maryland 20742, US}
\affiliation{ Institute for Physical Science and Technology, University of Maryland, College Park, MD 20742 USA}
\author{Jay Sau}
\affiliation{ Joint Quantum Institute and Condensed Matter Theory Center, Department of Physics,
University of Maryland, College Park, Maryland 20742, US}
\date{\today}

\begin{abstract}
Thouless pumping is a transport phenomenon whereby an insulating, time-periodic Hamiltonian induces a robustly quantized transfer of charge, per driving cycle, in the quasi-adiabatic limit. In this work we apply the shortcuts to adiabaticity (STA) method of counterdiabatic (CD) driving to the Rice-Mele model, an archetypal model for Thouless pumping. We show that the charge pumped rapidly across each bond of our CD Rice-Mele model is robustly quantized and determined by a Chern number.  
The CD Hamiltonian generally involves long-range, hence non-local, hopping. Therefore we also derive an exact, local, nearest-neighbor CD Hamiltonian for our lattice system. 
We show that our nearest neighbor non-adiabatic protocol possesses the same topological robustness to onsite disorder as quasi-adiabatic Thouless pumps and is moreover surprisingly robust to onsite disorder and noise errors in implementing the CD protocol, for sufficiently rapid driving. The protocol also reproduces finite-temperature Thouless pump results away from the adiabatic limit. In addition, we develop a general method to produce non-trivial, finite-time quantized pumping starting from any Rice-Mele initial ground state using only nearest-neighbor hopping. 

\end{abstract}

\maketitle
\section{Introduction}\label{intro} Thouless pumps are insulating systems that produce dissipationless, robustly quantized charge transport~\cite{thouless1983quantization,niu1984quantised}. In its simplest form, a Thouless pump is driven by the cyclic-in-time modulation of a spatially periodic potential, generating a dc current in the absence of an external bias. 
The charge pumped per driving cycle is an integer multiple of an electron's charge.  The integer is the first Chern number, a topological invariant determined by the time-evolution of the system's occupied states. So long as the system remains an insulator, the charge pumped per cycle is invariant to weak perturbations, such as lattice disorder or finite-range inter-particle interactions.

In Thouless' original formulation, the system's evolution is quasi-adiabatic, i.e. slow relative to the time scale set by the many-body ground state gap. Slow driving, however, makes the system susceptible to decoherence effects and is cumbersome for quantum technology applications, such as computation or sensing~\cite{guery2019shortcuts}. To overcome these limitations, we show how to accelerate Thouless pumping to produce quantized, dissipationless and robust transport in finite time.

Adiabatic Thouless pumps now constitute the backbone of the modern theory of polarization~\cite{king1993theory,resta2007theory}.  They hold promise in metrology, as a means to define the Coulomb~\cite{citro2023thouless,niu1990towards,pekola2013single}; for nano-machines in the form of Thouless motors~\cite{qi2009field,bustos2013adiabatic} and quantum nanowires~\cite{qi2009field}; and for quantum computing~\cite{das2006controlled}. Beyond physics, Thouless pumps have recently been proposed as a method to greatly boost piezoelectricity in conjugated organic polymers~\cite{villani2024giant}, and to molecularly engineer cyclic transitions between the bonding and antibonding orbitals of trans-polyacetylene via Floquet-Thouless pumping~\cite{zhou2021first,zhou2023molecular}.

Although Thouless pumps were predicted theoretically over forty years ago, only in the past decade have they been realized experimentally in extended, many-body $1$d closed systems -- the setting originally considered by Thouless. In the 1990's adiabatic quantum-dot (QD) pumps were realized as open $0$d systems~\cite{keller1996accuracy,switkes1999adiabatic,pekola2008hybrid}. QD pumps rely on arbitrarily small transmission through the barriers separating the QD from the leads, which requires a choice of large barrier height. In contrast, zero transmission for any barrier height is automatically achieved by replacing the QD with a Thouless pump between the leads in the thermodynamic limit, due to the gap of a Thouless pump system. Since 2016, closed, $1$d, {\it adiabatic} Thouless pumps based on the Rice-Mele model~\cite{rice1982elementary} have been observed in ultracold atom platforms~\cite{nakajima2016topological,lohse2016thouless, citro2023thouless}, including implementations with lattice disorder~\cite{nakajima2021competition}, particle-particle interactions~\cite{walter2023quantization}, and as open systems~\cite{fedorova2020observation}. They have also been realized in superconducting qubits~\cite{liu2025interplay} and qutrits~\cite{tao2025emulating}, spin systems~\cite{schweizer2016spin}, and photonic systems~\cite{nakajima2021competition,cerjan2020thouless}, the latter including those modeling lattice disorder~\cite{jurgensen2023quantized}. All of these implementations have been nearest-neighbor models. Adiabatic, nearest-neighbor Thouless pumps have also been realized in magneto-mechanical systems~\cite{grinberg2020robust} including with disorder~\cite{jurgensen2025quantized}, as well as in photonic crystal platforms~\cite{kraus2012topological,ke2016topological,verbin2015topological,cerjan2020thouless}, again including systems with particle-particle interactions~\cite{tangpanitanon2016topological,jurgensen2021quantized}. 

A growing field of study aims to accelerate topological charge pumping. One approach treats Thouless' time-periodic Hamiltonian as a Floquet Hamiltonian. In this case, states in a gapped Floquet band structure can exhibit quantized charge pumping~\cite{kitagawa2010topological}. Similar non-adiabatic charge pumping can occur in the $2 d$ anomalous Floquet-Anderson insulator (AFAI)~\cite{titum2016anomalous}. 
However, in the $1d$ Floquet-Thouless pump, quantized pumping requires a fine-tuned initialization of the system to occupy a quasi-energy band, which generally differs from the system's ground state. Indeed, if a Floquet drive is instead applied to the system initialized in its ground state, the quantization of charge pumping has corrections that scale as the square of the drive speed even after a finite number of cycles~\cite{privitera2018nonadiabatic}. In the AFAI, due to its additional spatial dimension, the filling of a particular Floquet band can be accomplished without fine tuning, by initializing the system with a fully filled fraction of states starting from one spatial edge. In the many-cycles/long-time limit, the system converges to a fully filled Floquet band and thus produces quantized charge pumping. Other related, non-adiabatic quantized transport effects are topological, quantized energy pumps via multiple incommensurate drives in many-body quasi-periodic systems~\cite{long2021nonadiabatic,nathan2021quasiperiodic}. Explicitly including damping can also lead to fast, quantized charge pumping in open dissipative systems via a non-Hermitian generalization of the Rice-Mele Hamiltonian~\cite{fedorova2020observation}.

All of the above methods proposed to accelerate topological pumps differ in essential aspects from Thouless' original setting, in which a closed, $1d$ many-body quantum system, initialized in its ground state, driven around one cycle, produces robustly quantized, dissipationless, pumping while remaining approximately in the ground state.

Shortcuts to adiabaticity (STA) offer a quantum control-based line of attack to accelerate transport while maintaining a system in its ground state~\cite{guery2019shortcuts}. In the counterdiabatic (CD) approach to STA, given a time-dependent Hamiltonian $H_0(t)$ and a system initialized in one of its instantaneous eigenstates, adding an appropriately designed auxiliary term $H_{CD}(t)$ causes the system to remain in this eigenstate of $H_0(t)$ at all times. The CD Hamiltonian $H_{CD}(t)$ suppresses transitions to other eigenstates~\cite{demirplak2003adiabatic,demirplak2005assisted,berry2009transitionless}. STA methods have been been realized experimentally in cold atoms~\cite{Schaff10,Schaff11}, nitrogen-vacancy (NV) centers~\cite{Kolbl19}, trapped ions~\cite{Kihwan16}, superconducting circuits~\cite{lu2017nonleaky}, and quantum dots~\cite{Liu24}; see~\cite{guery2019shortcuts} for a comprehensive review. Their applications include the cooling of optomechanical resonators~\cite{LiWuWang11}, the rapid loading of Bose-Einstein condensates in optical lattices~\cite{Masuda14,Zhou-NJP18}, and the implementation of fast quantum gates in NV centers~\cite{Song16} and superconducting qubits~\cite{Wang18}.
CD methods have been proposed to speed up both classical~\cite{funo2020shortcuts,takahashi2020nonadiabatic} and $0d$ quantum geometric Cooper-pair pumps~\cite{erdman2019fast}, as well as single-body, quasi-periodic topological energy pumps~\cite{crowley2019topological}.
To our knowledge, the first examples of STA methods applied to Thouless pumps are in the works of~\cite{malikis2022ideal,schouten2021rapid} which construct the Lewis-Riesenfeld invariants for a highly constrained class of clean, Rice-Mele models that can be exactly mapped onto the Sinh-Gordon equation. They use this mapping and the invariant they construct to prove the existence of finite frequency Floquet eigenstates with non-trivial winding numbers for the class of RM models they consider.

In this work, we apply the STA method of CD driving, to Thouless' topological pump: a many-body, non-interacting $1d$ fermionic chain, initialized in the ground state \cite{thouless1982quantized} and driven around a single cycle. We focus on the Rice Mele (RM) model~\cite{rice1982elementary}, whose Hamiltonian $H_0(t)$ is given by Eq.~\eqref{eq:RMH-space}. The system evolves under the Hamiltonian $H_0(t) + H_{CD}(t)$ as both $H_0(t)$ and $H_{CD}(t)$ vary periodically in time for one period. By adding $H_{CD}$ to $H_0$, we preserve the filled single-particle Bloch states of $H_0$ throughout the system's evolution. We analytically show that under CD driving, the charge pumped across each bond of the system is given by the Chern number derived from the dynamics of the ground state density matrix of $H_0$. Thus for all pump cycles for which $H_0$ has a non-vanishing gap, the corresponding CD pumped charge is robustly quantized as in Thouless' original construction.

While CD driving produces quantized, arbitrarily rapid particle transport, the required $H_{CD}$ generally contains long-range hopping terms. Engineering such long range hopping terms requires implementations using the Floquet-Magnus expansion or quantum computing circuits as opposed to local hopping that can realized with simple periodic potentials. Indeed, a recent subbranch of STA research focuses on developing local approximate control protocols in many-body systems~\cite{claeys2019floquet,vcepaite2023counterdiabatic,sels2017minimizing,kolodrubetz2017geometry,Gjonbalaj_2022,morawetz2024efficient}. In particular, the work of~\cite{sels2017minimizing}, considers a $1$d lattice model of free fermions, and argues that the generic exact CD term is non-local, in that it involves long range hopping. As a result, they construct an approximate CD operator that is nearest-neighbor-local by construction. We take a different track to obtain an exact, analytic, nearest-neighbor-local protocol to {\it accelerate} Thouless pumping, inspired by the ``control-freak" limit~\cite{asboth2016short} of the RM model, which we refer to as the ``counterdiabatic bucket brigade protocol" (BBCD). We also demonstrate the generality of finding exact, nearest-neighbor, counterdiabatic RM pumping protocols: using only nearest neighbor hopping, and numerical optimization, we can drive a system initialized in the ground state of an arbitrary RM Hamiltonian, to the ground state of our bucket brigade Hamiltonian. We can thus achieve finite-time, topologically non-trivial, quantized pumping with a nearest-neighbor Hamiltonian starting from any initial RM model. 

Moreover, we show that our BBCD protocol reproduces standard quasi-adiabatic Thouless pump results for weak disorder and finite temperature, away from the adiabatic regime. We also illustrate a general feature of CD driving: dynamic robustness. For a system initialized in an eigenstate of $H_0$, for sufficiently fast driving, the presence of strong noise and disorder to the total $H$ do not impact the state transfer. Here, strong means relative to gap of $H_0$. Thus, the pumped charge remains quantized far from the adiabatic limit, even in the presence of strong noise and disorder. Therefore, our BBCD protocol possess two types of robustness away from the adiabatic limit: the topological, or gap-of-$H_0$-based robustness as in standard Thouless pumps, as well as a dynamic robustness due to CD driving.

\section{CD Rice-Mele Model}\label{RMmodel} 
\begin{figure}[ht]
\includegraphics[width=\linewidth]{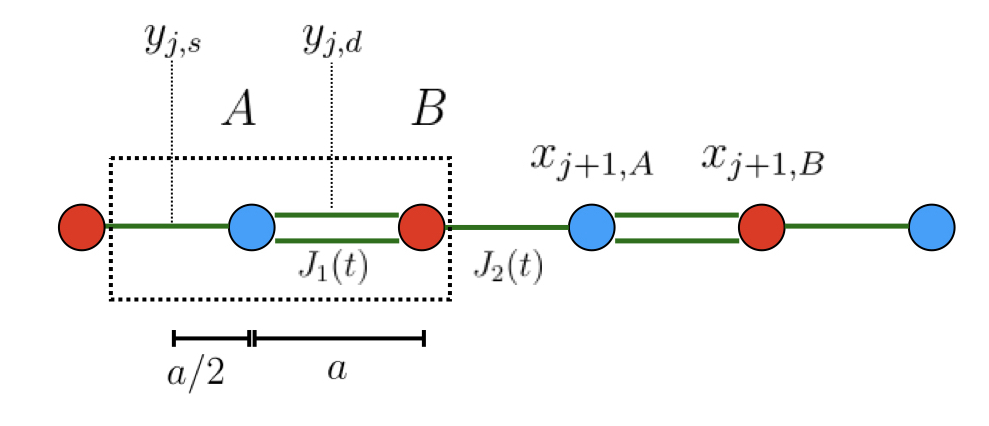}
\caption{\textbf{The Rice-Mele model} consists of a $1$d lattice of two atomic species, $A$ and $B$, at positions $x_{j,A}=2aj$ (blue) and $x_{j,B}= a(2j+1)$ (red), connected by intracell bonds (d, double green lines) at positions $y_{j,d} = a(2j+1/2)$ and intercell bonds (s, single green lines) at positions $y_{j,s} = a(2j-1/2)$. The dashed box shows a unit cell of length $2a$. The intracell/intercell hopping amplitudes are $J_{1/2}$}
\label{fig:RM}
\centering
\end{figure}

The Rice-Mele model provides a useful lattice implementation of non-interacting Thouless pumping. The model consists of a $1d$ chain of atoms, two per unit cell, with both intracell and intercell bonds; see Figure \ref{fig:RM}.

There are $N$ unit cells, and we impose periodic boundary conditions.
In real space, the RM Hamiltonian is
\begin{eqnarray}
\label{eq:RMH-space}
    \hat{H}_0(t) &=& -\sum_j \left( J_1(t)\hat{b}_j^{\dagger}\hat{a}_j + J_2(t)\hat{a}_{j+1}^{\dagger}\hat{b}_j + h.c.\right) \nonumber\\
    &-& \Delta(t)\sum_j\left(\hat{a}_j^{\dagger}\hat{a}_j - \hat{b}_j^{\dagger}\hat{b}_j\right)
\end{eqnarray}
where $\hat{a}_j^{\dagger}(\hat{a}_j)$ and $\hat{b}_j^{\dagger}(\hat{b}_j)$ are creation (annihilation) operators acting on the left and right sites of the $j$th unit cell, of cell length $2a$. The coefficients $J_1(t)$, $J_2(t)$ and $\Delta(t)$ tune the intracell hopping amplitude, intercell hopping amplitude, and a staggered on-site potential respectively.
We take $J_{1/2}(t)$ and $\Delta(t)$ to be periodic on the interval $[0,T]$, and constant outside the interval.

To write the RM Hamiltonian in momentum space, let $\tilde{X}_k \equiv N^{-1/2} \sum_j \hat{X}_j e^{-i2ajk}$, with $\hat X=\hat a, \hat b$, denote the Fourier transform of the orbital operator $\hat{a}_j$ or $\hat{b}_j$.
The wave number takes on values $k=\pi n/aN$, where $-N/2\le n < N/2$ and $n\in\mathbb{Z}$, where $a$ is the lattice constant.
Then
\begin{align}\label{RMHfull}
\hat{H}_0(t) = \sum_k 
        \begin{bmatrix}
           \tilde{a}_k^\dagger \, \,\tilde{b}_k^\dagger
         \end{bmatrix} \hat{H}_0(k,t) \begin{bmatrix}
           \tilde{a}_k \\
           \tilde{b}_k
         \end{bmatrix} \, ,
\end{align}
with
\begin{align}\label{RMH}
    \hat{H}_0(k,t) = \bm{R}(k,t)\cdot\bm{\sigma}.
\end{align}
Here, $\bm{\sigma}=(\sigma_x,\sigma_y,\sigma_z)$ is a vector of Pauli matrices acting on the space of Bloch states $[\tilde{a}_k^\dagger,\tilde{b}_k^\dagger]^T$, and  
\begin{equation}\label{RMvec}
\bm{R}(k,t) = \begin{bmatrix}
           -J_1(t)-J_2(t)\cos(2ka) \\
           -J_2(t)\sin(2ka) \\
           -\Delta(t)
         \end{bmatrix}.
\end{equation}
$\hat H_0(k,t)$ is gapped, unless $J_1(t) - J_2(t) = \Delta(t) = 0$. 

Eq.~\eqref{RMH} has the same form as that considered by Berry~\cite{berry2009transitionless} (see Eq.\,3.1 therein), who solved for $\hat H_{CD}$ for a spin-1/2 particle in a magnetic field. Since Eq.~\eqref{RMH} describes a two-band system, we invoke Berry's result (Eq. 3.8 of Ref.~\cite{berry2009transitionless}) to obtain $\hat H_{CD}(k,t)$ in lattice-momentum space:
\begin{equation}\label{Htot}
\begin{split}
\hat{H}(k,t) &= \hat{H}_0(k,t)+\hat{H}_{CD}(k,t) 
\\
&= \left(\bm{R}+\frac{\hbar}{2R^2}\bm{R}\times\partial_t\bm{R}\right)\cdot\bm{\sigma} \\
&\equiv \bm{u}(k,t)\cdot\bm{\sigma}.
\end{split}
\end{equation}

If a system begins in the single-particle ground Bloch state of $\hat H_0(t=0)$, and then evolves under
\begin{align}\label{Htotk}
\hat{H}(t) = \sum_k 
        \begin{bmatrix}
           \tilde{a}_k^\dagger \, \,\tilde{b}_k^\dagger
         \end{bmatrix} \hat{H}(k,t) \begin{bmatrix}
           \tilde{a}_k \\
           \tilde{b}_k
         \end{bmatrix} \, ,
\end{align}
with $\hat H(k,t)$ given by Eq.~\eqref{Htot}, then it remains at all times in the single-particle ground Bloch state of $\hat H_0(t)$, as in Thouless' original setting~\cite{thouless1982quantized} (rather than in a Floquet-Bloch state, as in Ref.~\cite{kitagawa2010topological}), under arbitrarily fast driving. For this system, we will show that the particle transport is quantized. 

\section{Non-Adiabatic Quantization of Pumped Charge}\label{pumpedcharge} Since Eq.~\eqref{Htot} is not a kinetic-plus-potential Hamiltonian, we use a generalized Peierls substitution method~\cite{bernevig2013topological} to define a current operator from which we can extract the physical current. For a translationally invariant lattice with basis, we find the average current to be:
\begin{equation}
\begin{split}\label{unit_cell_cur}
    j_{avg}(t)= \frac{e}{\hbar}\int \frac{dk}{2\pi}\left(\partial_k\bm{u}\cdot\hat{\bm{R}}-\hbar\partial_t{\rm Tr}\left[\hat{c}\hat{\rho}_k^{\dagger}\right]\right),
    \end{split}
\end{equation}
where $\hat{\rho}_k(t) = \frac{1}{2}\left(\mathbb{I} - \hat{\bm{R}}(k,t)\cdot\bm{\sigma}\right)$ is the single-particle, ground-state density matrix, $\hat{c}\,\,  =\sum_{c\in \{0,a\}}c\ket{c}\bra{c}$, is the position operator projected onto basis sites within a unit cell, and $\hat{\bm{R}} = \bm{R}/R$. See Appendix~\ref{appendix:current} for details.

The average charge pumped across a unit cell is the time-integral of the current in the unit cell, averaged over the cell's bonds:
\begin{align}\label{Qt}
    &Q_{pump}(t)=\int_0^t d\tau \, j_{avg}(\tau),
\end{align}
Integrating~\eqref{Qt} by parts, and substituting $\bm{u}$ from Eq.\eqref{Htot} we find 
\begin{equation}\label{eqQpumpC}
\begin{split}
       Q_{pump}(t)&=e\biggl(\frac{1}{4\pi}\int_0^t d\tau\int_{-\frac{\pi}{2a}}^{\frac{\pi}{2a}}dk \, \hat{\bm{R}}\cdot\left(\partial_k\hat{\bm{R}}\times\partial_t\hat{\bm{R}}\right)
       \\
       &+\frac{1}{2\pi}\int_{-\frac{\pi}{2a}}^{\frac{\pi}{2a}}dk\left(\frac{a}{2}\right)(\hat{\bm{R}}_z(k,t)-\hat{\bm{R}}_z(k,0))\biggr).
\end{split}
\end{equation}
where $\hat{\bm{R}}_z$ is the $z$th component of $\hat{\bm{R}}$. Setting $t=T$, the first term in parentheses on the right is the Chern number $C$ of the vector field $\hat{\bm{R}}(k,t)$ over the $(k,t)$-torus. $C$ is an integer as long as $\hat{\bm{R}}(k,t)$ is a continuous vector field on the torus due to $\hat{\bm{R}}$'s periodic boundary conditions in $k$ and $t$~\cite{hasan2010colloquium}.
The second term vanishes when $t=T$, by the time-periodicity of $\hat{\bm{R}}$. Thus, the charge pumped over one cycle, $Q_{pump}(T)$, is an integer multiple of $e$.

$Q_{pump}(t)$ is the pumped charge averaged over the bonds per unit cell, and does not provide a full picture of the flow of charge across individual sites, nor across bonds within (between) the unit cell(s). The former information is provided by the dynamics of the time-dependent charge on site $x$, $Q(x,t)\equiv e{\rm Tr}(\hat{\rho}(t)\ket{x}\bra{x})$ (so $Q(x,t) >0$ for an occupied site), from which we obtain 
\begin{equation}\label{eqQ}
\begin{split}
    Q(c,t)&=e\frac{a}{\pi}\int_{\frac{-\pi}{2a}}^\frac{\pi}{2a}dk \rho_k(c,c,t)
    \\
    &=e \frac{a}{2\pi}\int dk (1-\hat{\bm{R}}_z(k,t)s(c)) .
\end{split}
\end{equation}
Here, $c \in \{0,a\}$, $\rho_k(c,c,t)$ are the diagonal elements of the $2\times2$ unit-cell density matrix, and $s(c) \in \{-1,1\} $. For a given cell, we take the $A$ and $B$ atoms to be located at $c = 0$ and $a$ and the single and double bonds' locations to be $y_s = -a/2$ and $y_d = a/2$, respectively. From Eqs.~\eqref{RMHfull} and~\eqref{RMH} we obtain $s(0)= (\sigma_z)_{11} = 1$ and $s(a) = (\sigma_z)_{22} =-1$. 
The charge pumped across any bond $b\in\{s,d\}$ is $Q_{pump,b}(t) = \int d\tau \langle J(y_b)\rangle$, where 
\begin{equation}\label{eq:exv_currentBop}
    \langle J(y_b)\rangle = {\rm Tr}(\hat{J}(y_b)\hat{\rho})
\end{equation} is the expectation value of the current operator that we derive in Eq.~\eqref{eq:Jreal} in Appendix \ref{appendix:current}. From this current operator we can derive the conservation law $\partial_t Q(x,t)=-J(x+a/2,t)+
J(x-a/2,t)$ and $Q_{pump}(t) = \frac{1}{2}\left(Q_{pump,s}(t)+Q_{pump,d}(t)\right)$, from which we then find
\begin{align}\label{qpumpb}
\begin{split}
    Q_{pump,b}(t)&=Q_{pump}(t)
    \\
    &-\frac{\text{sign}(y_b)}{2}(Q(0,t)-Q(0,0)).
\end{split}
\end{align}
This result captures the rearrangement of charge within the unit cell (setting $b=d$), and the charge pumped between unit cells ($b=s$). Note that the second term vanishes at $t = T$, and thus the charge pumped across any bond over a cycle is also a Chern number.
\begin{figure*}
\includegraphics[width=0.8\textwidth]{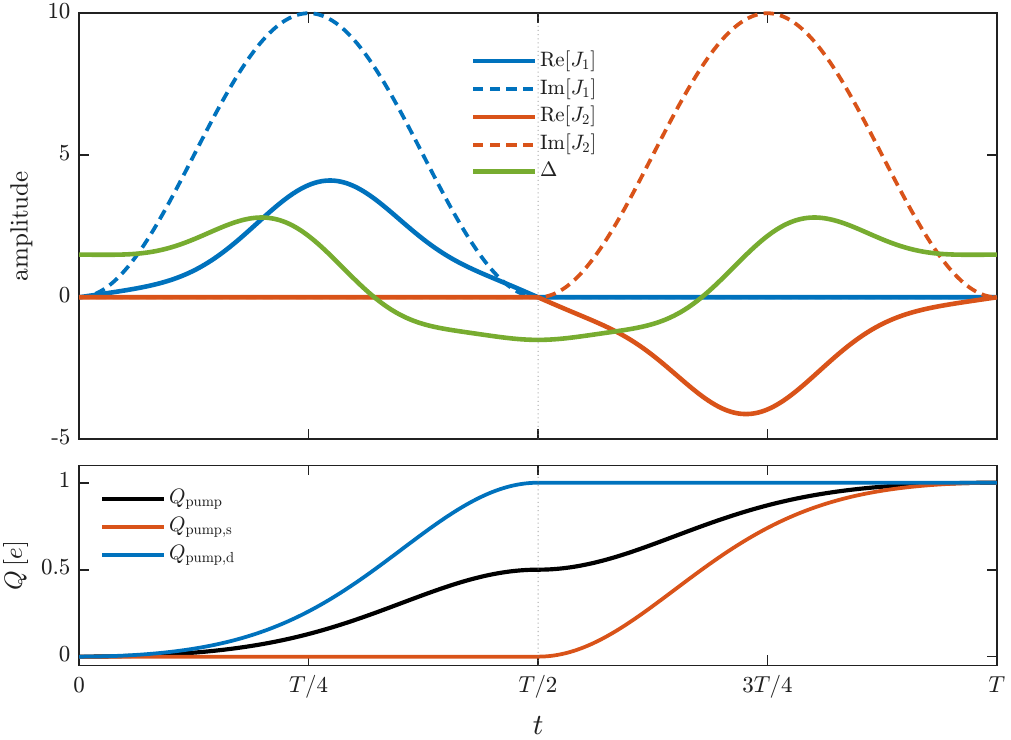} 
\caption{\textbf{Bucket brigade CD Thouless pump:} Real space hopping and potential amplitudes for the nearest-neighbor bucket brigade counterdiabatic protocol. We choose $\theta(t) = \omega t -(1/2)\sin(2\omega t)$, where $\omega = 2\pi/T$ determines the speed of the drive and $\lambda(t) =1.5 +3\sin^4(2\omega t)$ which sets the gap of $H_0$; these satisfy the boundary conditions described in the main text}
\label{fig:RealSpace}
\end{figure*}

\section{Bucket Brigade Counterdiabatic (BBCD) protocol}\label{sec:cleanBBCD} 
We derive an analytically soluble, topologically non-trivial, limit of counterdiabatic Thouless pumping that is inspired by the so-called ``control freak" protocol~\cite{asboth2016short}. This protocol produces quantized pumping in the quasi-adiabatic limit by requiring one of $J_{1,2}(t)$ in Fig.~\ref{fig:RM} to vanish at any instant of time, so that the lattice is never completely connected, while using intracell bias terms to ``bucket brigade" the particles along the chain. Inspired by this physical picture, along with CD driving, we produce an analytic finite-time protocol with quantized pumping. We use the following protocol structure:  
\begin{equation}\label{eq:Rex}
\begin{split}
    &\bm{R}(k,0<t<T/2)=(-R_\perp(t),0,-R_{||}(t))
    \\
    &\bm{R}(k,T/2<t<T)=
    \\&(-R_\perp(t)\cos(2ak),-R_\perp(t)\sin(2ak),-R_{||}(t)).
    \end{split}
\end{equation}
At the start of the first phase, $0<t<T/2$, electrons are delocalized between the A and B sites within the unit cell, across the $J_1$ bonds. At the start of the second phase, $T/2<t<T$, electrons are delocalized across a B site of one unit cell and an A site of the next unit cell, across the $J_2$ bonds. The former set of bonds are entirely inside the unit cell (i.e. double bonds in Fig.~\ref{fig:RM}) and therefore lead to a Hamiltonian with no $k-$dependence. Substituting Eq.~\eqref{eq:Rex} into Eq.~\eqref{Htot}, the total Hamiltonian to implement this protocol is given by a vector \begin{equation}
\begin{split}\label{eq:uCDk}
&\bm{u}(k,0<t<T/2)=\bm{R}(k,t)+(\hbar\dot{\theta}/2)(0,1,0),
\\
&\bm{u}(k,T/2<t<T)=
\\
&\bm{R}(k,t)+(\hbar\dot{\theta}/2)(-\sin{(2ak)},\cos{(2ak)},0).
\end{split}
\end{equation}
We have reparameterized the time-dependence of  $R_{\perp,||}(t)$ as $R_{||}(t)=e^{\lambda(t)}\cos{\theta(t)}$ and  $R_{\perp}(t)=e^{\lambda(t)}\sin{\theta(t)}$. Continuity and time-periodicity of $\bm{u}(k,t)$ require $\dot{\theta}(t=0) = \dot{\theta}(T/2)=\dot{\theta}(T) = 0$. Continuity and time-periodicity of the vector $\bm{R}(k,t)$ require  $\theta(t=0)=0$, $\theta(T/2)=\pi$ and $\theta(t=T)=2\pi$. With these choices, this protocol corresponds to a five-step physical sequence, whose interpretation becomes clear upon comparing to Eq.~\eqref{RMvec}. At $t=0$, the system begins as a disconnected monomer chain. Then during $0<t<T/2$, it is a topologically trivial Su-Schrieffer-Heeger (SSH) chain ($J_1 = R_{\perp} > J_2 = 0$). At $t = T/2$ it is again a disconnected monomer chain. During $T/2<t<T$ it is a topological SSH chain ($0 = J_1 < J_2 = R_{\perp}$), and lastly, at $t=T$ is again a disconnected monomer chain. These are essentially the same physical steps as the control freak protocol~\cite{asboth2016short}, except that in place of monomer steps over intervals of $t$, here, these steps occur instantaneously at $t = 0,\,T/2,\, \text{and } T$. The resulting bucket brigade counterdiabatic protocol hopping and potential terms are plotted in Fig.~\ref{fig:RealSpace}. The real-space hopping amplitudes have an imaginary component, which can be removed from the Hamiltonian by rotating the $A$ lattice relative to the $B$ lattice with a time-dependent phase: $\hat{a}'_{j}(t) = \exp\left(-i \tan^{-1}\left(\frac{\hbar\dot{\theta}}{2\sin(\theta)}\right)\right)\hat{a}_{j}$. This rotation does not affect the pumped charge which is proportional to the Chern density, which in turn is proportional to the polarization, given by the sum of terms of the form $\hat{a}'^{\dagger}_j\hat{a}'_j-\hat{b}^{\dagger}_j\hat{b}_j = \hat{a}^{\dagger}_j\hat{a}_j-\hat{b}^{\dagger}_j\hat{b}_j$. Substituting $\bm{R}$ from Eq.~\eqref{eq:Rex} into the pumped charge Eq.~\eqref{eqQpumpC}, and the charge on site $x$, Eq.~\eqref{eqQ}, we find 
\begin{equation}
\begin{split}\label{Qavgcontrol}
Q_{pump}(t)&=\frac{e}{4}\biggl(1-\cos(\theta(t))
\\
&+2\left(1+\cos(\theta(t))\right)\Theta\left(t-\frac{T}{2}\right)\biggr)
\end{split}
\end{equation} and $Q(0,t)-Q(0,0)=\frac{e}{2}(\cos{\theta(t)-1})$ and $\Theta(t)$ is the Heaviside step function. These results can be combined, using Eq.~\eqref{qpumpb}, into equations for charge pumped across each bond:

\begin{equation}\label{Qcontrol}
\begin{split}
    Q_{pump,s}(t) &= \frac{e}{2}\left(1+\cos(\theta(t))\right)\Theta\left(t-\frac{T}{2}\right), 
    \\
    Q_{pump,d}(t)&= \frac{e}{2}\biggl(1-\cos(\theta(t))
    \\
    &+\left(1+\cos(\theta(t))\right)\Theta\left(t-\frac{T}{2}\right)\biggr),
\end{split}
\end{equation}
where $\theta(t)$ varies from $0$ to $2\pi$ over the protocol. 
No charge is pumped across the $J_1$ ($s$) bonds during the first phase, or across the $J_2$ ($d$) bonds during the second phase. From the boundary conditions for $\theta(t)$ it is clear that both $Q_{pump,s}$ and $Q_{pump,d}$ increase from $0$ to $1$ over a period $T$, see the lower panel of Fig~\ref{fig:RealSpace}.

\begin{figure}[!t]
    \centering
    \begin{subfigure}[b]{0.98\linewidth}
        \centering
        \includegraphics[width=0.98\columnwidth]%
            {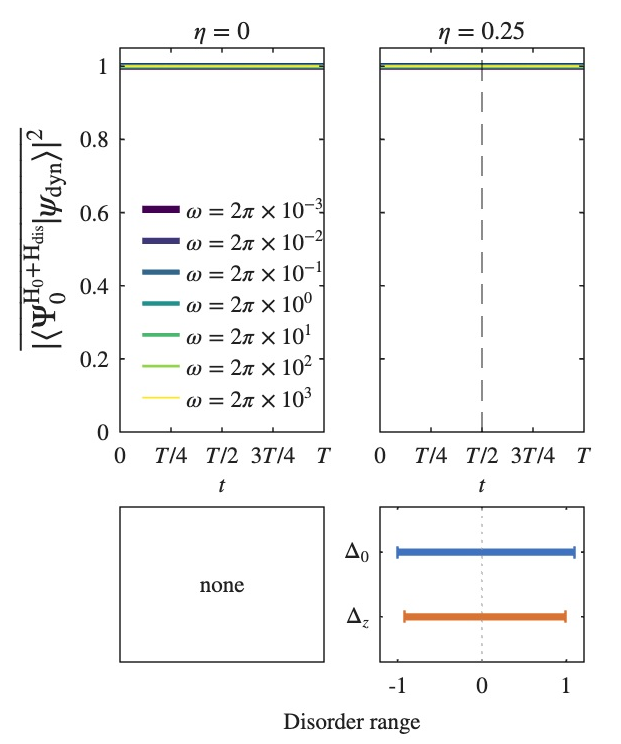}
        \caption{}
        \label{fig:RobustDisCDov}
    \end{subfigure}

    \vspace{0.4em}

    \begin{subfigure}[b]{0.94\linewidth}
        \centering
        \includegraphics[width=0.94\columnwidth]%
            {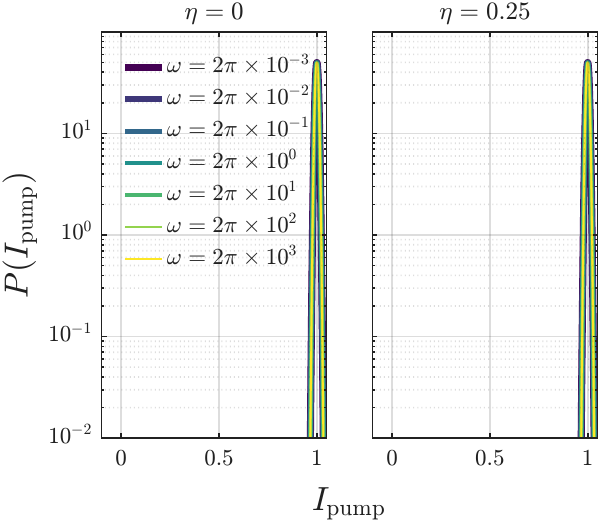}
        \caption{}
        \label{fig:RobustDisCDPC}
    \end{subfigure}

    \caption{\textbf{State transfer and pumped charge under known-disorder BBCD driving} (a) Mean overlap of the ground-state of $H = H_0 + H_{dis}$ with the dynamic state produced by $H+H_{CD}[H]$ (i.e. disorder-aware CD driving) versus time, with a sweep through seven orders of magnitude of driving speed for both the clean and weak disorder systems, where $\eta$ is an overall scaling strength for the disorder. The
    realized disorder ranges are below each panel. (b) The pumped-charge
    distributions $P(I_{\mathrm{pump}})$, which we see are delta functions centered on $1$. Plots produced for $N=100$ cells with P.B.C. averaged over $100$ realizations of the disorder}
    \label{fig:RobustDisCD}
\end{figure}

\begin{figure*}[h!t]
    \centering

    \begin{subfigure}{\textwidth}
        \centering
        \includegraphics[width=0.98\textwidth]{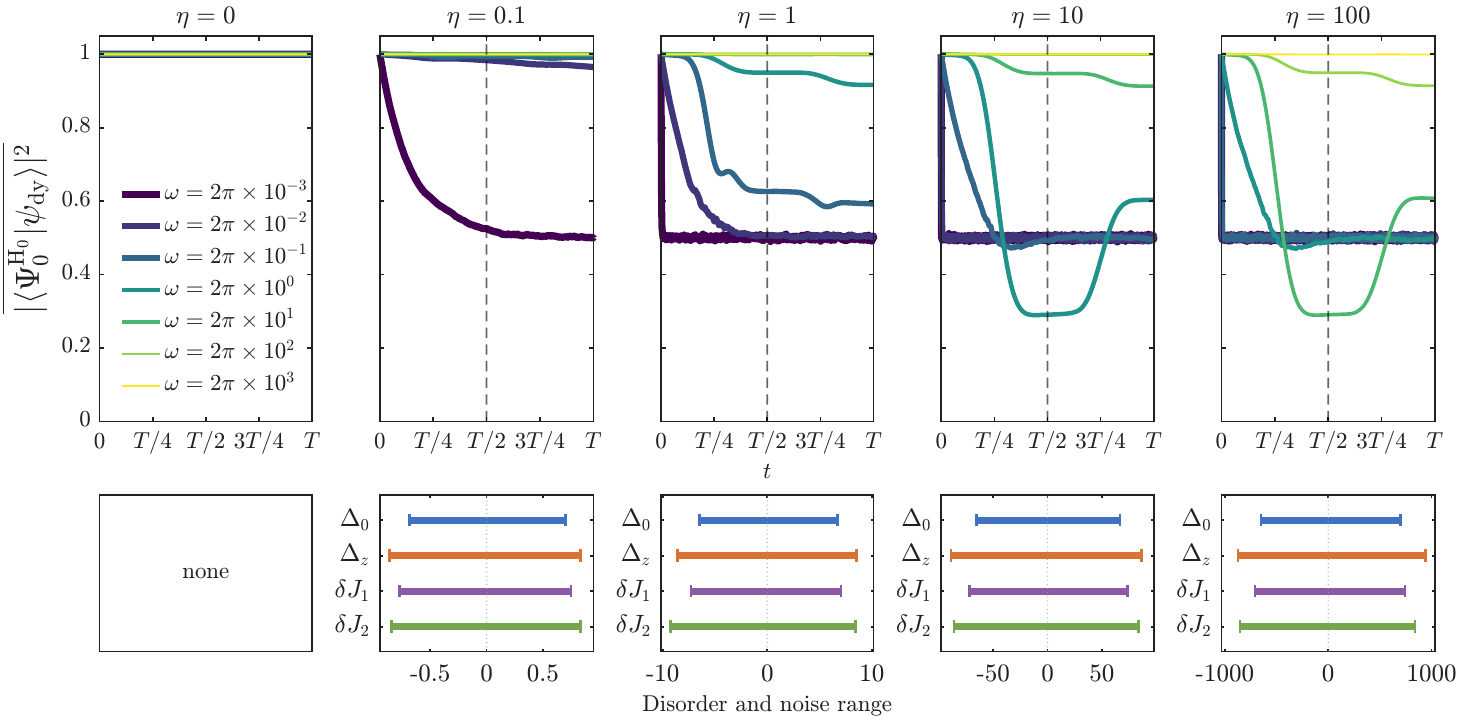}
        \caption{}
        \label{fig:RobustTrip_a}
    \end{subfigure}

    \vspace{0.5em} 

    \begin{subfigure}{\textwidth}
        \centering
        \includegraphics[width=0.98\textwidth]{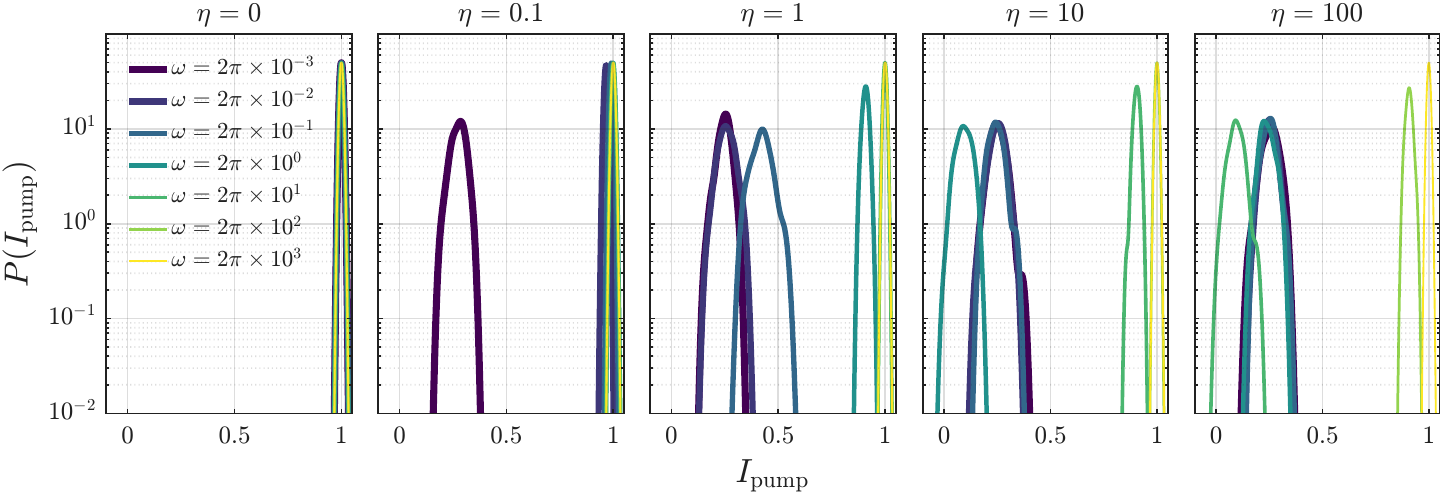}
        \caption{}
        \label{fig:RobustTrip_b}
    \end{subfigure}

    \caption{\textbf{State transfer and pumped charge under imperfect BBCD driving subject to unknown disorder and noise errors} Near-perfect state transfer and pumped charge result when the driving speed of the BBCD protocol exceeds the time scales set by the disorder and noise by an order of magnitude, which are greater than or equal to the size of $H_0$'s gap (order $1$). \textbf{(a)} Mean overlap of the ground-state of $H_0$ with the dynamic state produced by $H_0(t) + H_{dis}(t)+H_{CD}[H_0]$ (disorder and noise blind CD driving) versus time, with a sweep through seven orders of magnitude of driving speed, and a range of onsite disorder and onsite and hopping noise strengths, where $\eta$ is a  scaling strength to the disorder and noise. The realized disorder and noise ranges are below each overlap plot. $H_{dis}(t)$ is defined in Eq.~\eqref{eq:timeDisNoise}. \textbf{(b)} A log-scale plot of the corresponding pumped charge probably distribution $P(I_{pump})$ as a function of the $I_{pump}$. The pumped charge distribution is a delta function centered on $I_{pump} =1$ for all speeds in the clean system, and also for cycles that are short relative to the disorder and noise strength. For all other cases, the distribution is Gaussian-like, reflecting the breakdown of perfect quantization of pumped charge. Plots produced for a $N = 100$ dimer RM chain, with P.B.C. averaged over $100$ realizations of the disorder and noise}
    \label{fig:RobustTrip}

\end{figure*}
{\section{Topological Robustness of State Transfer and Pumped Charge}Away from the adiabatic limit there are two distinct forms of robustness: topological and dynamic. The first is a gap-based robustness to known, weak disorder analogous to adiabatic Thouless pumping. The second is a dynamical robustness to errors in implementing the correct CD protocol. We develop two analytic arguments, one here and one in Appendix~\ref{appendix:robustness_known_dis}, as well as provide numerical demonstration of the topological robustness of the BBCD protocol away from the adiabatic limit. We add onsite disorder to the bucket brigade (BB) Hamiltonian: $H_0(t) \to H_0(t)+H_{dis}$ with 
\begin{equation}\label{eq:DisHam}
    \begin{split}
    H_{dis}&=
        -\sum_{j=1}^{N}\eta\Biggl(d^{z}\Delta^{z}_j\ket{j}\bra{j}\otimes\sigma_z
       \\
       &+d^{0}\Delta^{0}_j\ket{j}\bra{j}\otimes\sigma_0\Biggl)
    \end{split}
\end{equation}
where $\Delta^{z}_i,\Delta_i^0$ are time-independent, uncorrelated random numbers drawn from the Gaussian distribution $N(0,1)$, with mean zero, and standard deviation one, and $d^{z},d^{0}$ are the disorder strengths. We assume the disorder is weak: $\eta d^z,\eta d^0 \ll e^{\lambda(t)}$, which is the gap of the BB $H_0(t)$. As in adiabatic Thouless pumping, we initialize our system in the ground state of $H_0(t) + H_{dis}$ by putting it in contact with a very low temperature ($\mathcal{T} \to 0$) thermal bath. We then disconnect the heat bath, resulting in a closed system initialized in its disordered ground state. Since the bucket-brigade protocol leads to a dimerized, or monomorized chain at all times, we again find the BBCD Hamiltonian is NN during the drive with a dimerization pattern that matches the BB $H_0$. As developed in Section~\ref{sec:cleanBBCD}, for the clean system, driven by the clean BBCD protocol, during the first leg of the drive, all $N$ of the $[A_i,B_i]$ dimers of the system are automatically synchronized in the dimer ground state of $(1,0)$ localized on the $A_i$ atom, and then are driven in-step within each dimer to $(0,1)$ to the state localized on the $B_i$ atom. During the second half of the drive, using the newfound dimerization of the $[B_{i},A_{i+1}]$ atoms, the dimers are again driven in step from $(1,0)$ on the $B_{i}$ atoms to $(0,1)$ on the $A_{i+1}$ atoms. So long as the disorder does not spoil the synchronization of the dimers at $t=0, T/2 \text{ and } T$, this same picture remains true for a range of disorder values up to a critical threshold. At these three values of $t$, $H_0$ is diagonal, and each dimer has a well-defined ground state; the synchronization criteria corresponds to ensuring that $H_0$ and $H_0 + H_{dis}$ have the same ground state for each dimer. The critical disorder threshold where the synchronization of the dimers breaks down is proportional to $H_0$'s gap at $t=0,T/2 \text{ and } T$, and for each leg of the protocol is given by 
\begin{equation}\label{eq:sync_cond}
\begin{split}
        -e^{\lambda(0)} = -\Delta(0) &\leq \eta d^z\Delta^{z}_j\leq \Delta(T/2) = e^{\lambda(T/2)}
        \\
        -e^{\lambda(T)}=-\Delta(T) &\leq D_{j,j+1}\leq \Delta(T/2)=e^{\lambda(T/2)}.
\end{split}
\end{equation}
For the dimers to remain synchronized, the above inequalities must hold for all $N$ dimers. 
So long as the states of the $N$ dimers remain synchronized, the BBCD protocol calculated for the disordered Hamiltonian, (see  Appendix~\ref{appendix:robustness_known_dis}) will again shuttle each dimer state from A to B and then to A in the next dimer. Thus, the pumped charge will remain exactly quantized just as in the clean case. We can demonstrate this robustness to disorder analytically using the dimerization of the chain under BBCD and the continuity equation per dimer (or, equivalently, the polarization per dimer), 
\begin{equation}
    \langle \hat{J}(t)\rangle = \frac{d}{dt}\left(\hat{Q}_{A}(t)-\hat{Q}_{B}(t)\right)
\end{equation} 
where the expectation value of the current operator is defined in Eq.~\ref{eq:exv_currentBop} (with bond argument suppressed) and $\hat{Q}_{i=A,B}(t)$ are the sublattice charge Heisenberg operators. We can compute the average pumped charge in the chain for the first half of the BBCD protocol through the $[A_i,B_i]$ dimers by integrating the continuity equation averaged over all of the system's dimers:
\begin{equation}\label{eq:dimerContEq1}
\begin{split}
   C_1 &= \frac{1}{2N}\int_0^{T/2} dt\sum_{j=1}^N \langle J(t)\rangle_{\psi_j}
   \\
   &=\sum_{j=1}^N\frac{\langle \hat{Q}_B(T/2)-\hat{Q}_B(0) - \hat{Q}_A(T/2)+\hat{Q}_A(0)\rangle_{\psi_j}}{2N}
   \\
   &=\frac{1}{2N}\sum_{j=1}^N\left(1-0 - 0+1\right) = 1
\end{split}
\end{equation}
where the expectation value is with respect to $\ket{\psi_j(t)}$, the state of the $j$th dimer at time $t$, which under CD driving is the instantaneous ground state of the dimer. The final line follows from the properties of the BB ground state when the synchronization criteria is satisfied: for all dimers, at $t=0$, the ground state is localized on the $A_i$ atoms, and at $t=T/2$ is localized on the $B_i$ atoms. Note, $e C_1 = Q_{pump,d}(T)$ as defined in Eq.~\eqref{qpumpb}. Similarly, for the second half of the BBCD protocol, using the newfound dimerization $[B_{i},A_{i+1}]$ we find
\begin{equation}\label{eq:dimerContEq2}
\begin{split}
    C_2 &= \frac{1}{2N}\int_{T/2}^{T} dt\sum_{j=1}^N \langle J(t)\rangle_{\psi_j}
   \\
   &=\sum_{j=1}^N\frac{\langle \hat{Q}_A(T)-\hat{Q}_A(T/2) - \hat{Q}_B(T)+\hat{Q}_B(T/2) \rangle_{\psi_j}}{2N}
   \\
   &=\frac{1}{2N}\sum_{j=1}^N(1-0-0+1) = 1
\end{split}
\end{equation}
where $eC_2 = Q_{pump,s}(T)$. Thus, as we found for the clean system in Section~\ref{sec:cleanBBCD}, the total, average pumped charge through the system is again $Q_{pump}(T) = \frac{1}{2}\left(Q_{pump,d}+Q_{pump,s}\right) = 1$. Fig.~\ref{fig:RobustDisCD} shows the comparison of applying the BBCD protocol in the clean system, as well as in presence of weak disorder to $H_0$ to a chain of $N =100$ dimers, for $100$ realizations of the disorder. We take $e^{\lambda(0)}=e^{\lambda(T/2)}=e^{\lambda(T)}=1$,
which also sets the scale of the adiabatic limit: $\omega_{ad} \ll 1$, where $\omega_{ad}$ is the adiabatic driving speed. Fig~\ref{fig:RobustDisCDov} shows the overlap of the instantaneous eigenstates of $H = H_0+H_{dis}$ with the dynamic state produced by $H_0 +H_{dis}+H_{CD}[H]$ which remains exactly $1$ over the course of the drive for both the clean system and the system in the presence of disorder: the BBCD protocol works exactly as advertised in the presence of weak disorder to $H_0$ that satisfies the synchronization condition. We write the CD term as $H_{CD}[H]$ to emphasize that it is computed using the eigenstates of the disordered Hamiltonian $H$. Moreover, Fig~\ref{fig:RobustDisCDPC} shows the the pumped charge distribution as well as the actual range of disorder values which we see satisfy the synchronization condition. Since we have disorder with zero-mean, the synchronization condition simplifies, and in the first leg is $\max_j(|\eta d_z\Delta^z_j|) = 0.99 < \min(\Delta(0),\Delta(T/2)) = 1$ and in the second is $\max_j(|D_{j,j+1}|) = 0.95 < \min(\Delta(T/2),\Delta(T)) = 1$. The pumped charge probably distributions $P(I_{pump})$ which are a function of pumped charge $I_{pump}$ are plotted using Kernel Density Estimation (KDE) with a bandwidth floor, on a logarithmic axis \cite{silverman2018density}. 

\section{Dynamic Robustness of State Transfer and Pumped Charge}We explore the dynamic form of robustness by introducing noise and disorder errors to the BBCD operator computed for the clean system $H_0$. We initialize the clean $H_0$ BB state by placing particles in alternating sites along a monomer chain, then turning on onsite potentials so that the occupied sites are at a lower energy than the unoccupied sites. We then drive with the BBCD protocol while also introducing unknown errors in the form of lattice disorder and noise in the potentials and hopping which the clean BBCD term does not take into account. Given that the CD operator is designed to exactly cancel all diabatic transitions one would expect CD driving to be fine-tuned and for the pumped charge quantization to break down under perturbations to $H = H_0 + H_{CD}[H_0]$. However, we find our BBCD pump possesses its own robustness to perturbations. We find that for a fixed disorder and noise strength -- even much larger than the gap of $H_0$ -- we can produce perfect quantization for sufficiently rapid driving. This seems remarkable given that $H_{CD}$ is designed to cancel transitions due to the non-adiabatic $H_0$ dynamics and not the dynamics generated by the noise and disorder. In Fig.~\ref{fig:RobustTrip} we plot the fidelity of the dynamical state with respect to the instantaneous ground state manifold of $H_0$, as well as the pumped charge distribution over $100$ realizations of the disorder and noise, for a range of noise and disorder strengths and driving speeds for a RM chain of $N = 100$ dimers. We see from Fig.~\ref{fig:RobustTrip_a} that if the disorder and noise are comparable to or greater than the size of $H_0$'s gap (order $1$), then driving speeds that are an order of magnitude larger than the time-scale set by the disorder and noise (each with its own energy scale) produce nearly perfect state transfer and pumped charge. Indeed, the pumped charge distribution is a delta function at $I_{pump}=1$ for all speeds for the clean system as well as for speeds that are fast relative to the disorder and noise strength time scales, and spreads into Gaussian-like distributions for intermediate-strong relative values of disorder and noise vs. driving speed, and for slow speeds and large disorder and noise it peaks far from $1$. We also note from Fig.~\ref{fig:RobustTrip_a} that for a fixed speed of driving the quantization breaks down with increasing noise and disorder strength. The details of the numerical implementation are in Appendix~\ref{appendix:robustness_unknown}. }

This dynamic robustness is due to the fact that for very fast driving, the large CD term acts over short driving period: if system is initialized in the correct ground state of $H_0(0)$, it will propagate to the corresponding ground state of $H_0(T)$ at time $T$ and thus produce quantized pumping; the faster the drive, the more robust the fidelity and quantization, as we see in Fig~\ref{fig:RobustTrip}.

We can demonstrate this CD-robustness analytically. Consider the limit of driving over extremely short intervals. To analyze this limit systematically, we write the Hamiltonian as $H = H(t/\epsilon)$, where $0 < \epsilon \ll 1$ and define a fast time variable $\tau = t/\epsilon$ where $\tau \in [0,T]$ which marks the advance of time during the driving interval $t \in [0,\,\epsilon \,T]$. We note that $\partial_t = \frac{1}{\epsilon}\partial_{\tau}$. Because the CD term contains a time derivative, we have $H_{CD}(\tau) = \frac{1}{\epsilon}\tilde{H}_{CD}(\tau)$ where $\tilde{H}_{CD}(\tau) = (\hbar/2)\bm{R}(\tau)\times\partial_{\tau}\bm{R}(\tau)/R(\tau)^2$. In the fast time variable, the CD Schrodinger equation is then
\begin{equation}\label{eq:scaling}
\begin{split}
    &\frac{i\hbar}{\epsilon}\partial_{\tau}\ket{\psi_n(\tau)} 
    \\
    &= \left(H_0(\tau) + \frac{1}{\epsilon}H_{CD}(\tau)+H_{dis}(\tau)\right)\ket{\psi_n(\tau)}.
\end{split}
\end{equation}
Multiplying through by $\epsilon$ and taking $\epsilon\to0$ shows that the dynamics are entirely determined by $H_{CD}$ which thus drives the system extremely rapidly from $\ket{\psi_n(0)}$ to $\ket{\psi_n(T)}$. This scaling argument holds for any $H_{CD}$, for any system and any perturbation that remains finite in the vanishing driving interval limit. In particular, it holds for any RM pumping protocol. Away from this limit, our numerical results demonstrate that for $\epsilon H_{dis} < 0.1$, we can produce near-perfect state transfer fidelity and quantization for a wide range of driving speeds and disorder and noise strengths. We note that for periodic boundary conditions, the qualitative  results of Figs.~\ref{fig:RobustDisCD} and~\ref{fig:RobustTrip} hold for $N=2$ dimers. For open boundary conditions, we find corrections to the pumped charge and state transfer, due to the diabatic transitions of edge states in the non-trivial SSH phase of RM, that vanish in the thermodynamic limit, where edge effects become exponentially suppressed in system size $N$.  

\section{Thermal Fluctuations}Regarding the impact of thermal fluctuations, we first note that the BBCD protocol has flat ($k$-independent) bands, whose gap is given by $\Delta E = E_1(t,k) - E_0(t,k) = 2|\bm{R}(t,k)| = 2e^{\lambda(t)}$, which thus determines the adiabatic limit. To explore the impact of finite temperature on the finite-time pumped charge, produced by the bucket-brigade CD protocol, we consider our system initially in equilibrium with a heat bath at temperature $\mathcal{T}$. The system's state is thus given by a Gibbs state under the Fermi-Dirac distribution. We then disconnect the heat bath, and evolve the system via unitary dynamics under the BBCD protocol. Thus, under CD driving the system's state is 
\begin{equation}
\rho(k,t) = \frac{1}{Z(k,0)}\sum_{n=0,1}e^{-E_n(k,0)/k_B\mathcal{T}}\rho_n(k,t)
\end{equation}
Because the BBCD protocol has flat bands, the phases are both constants as is $Z(k,0) = Z$, and thus the CD operator simply evolves $\rho_i(k,0)$ to time $t$ for $i = 1,2$. Thus, particle transport is given by a weighted sum of the transport produced by the ground and excited bands:
\begin{equation}
\begin{split}
    Q_{pump}^{th} &= \frac{1}{Z}\sum_{n=0,1}e^{-E_n(k,0)/k_B\mathcal{T}}Q_{pump}^{n}
    \\
    &=\tanh(e^{\lambda(0)}/k_B\mathcal{T})Q_{pump}^{0}
\end{split}
\end{equation}
The second line follows because the pumped charge per band is proportional to the Berry curvature $\Omega_i$ for each band $i = 0,1$, and because $\Omega_1 = -\Omega_0$. We thus see that the thermal contribution from the excited state transport is exponentially suppressed at low temperatures, and recovers the exactly quantized result in the temperature $\mathcal{T}\to0$ limit, just as in the adiabatic case~\cite{roura2025charge}. 

\section{Generic Nearest-Neighbor CD protocols}We now develop a method to construct generic, topologically non-trivial, nearest neighbor RM CD protocols based on the BBCD protocol. Our method works for an arbitrary RM model initialized in its ground state.
Equations~\eqref{Htot} and~\eqref{eqQpumpC} imply that if a Bloch Hamiltonian $H(k,t) = H_0(k,t)$ in the quasi-adiabatic regime gives rise to a Thouless pump, then $H(k,t) = H_0(k,t)+H_{CD}(k,t)$ gives rise to a Thouless pump for arbitrarily fast driving. Hence, to demonstrate a ``fast" Thouless pump, we can choose a form for $H(k,t)$ that is nearest-neighbor, and reverse-engineer a corresponding $H_0(k,t)$; if both are gapped, and periodic in $k$ and $t$, then $H$ and $H_0$ will produce finite-time and adiabatic Thouless pumping respectively, where the gap of $H_0$ determines the adiabatic limit.

The total Hamiltonian $H$ generated using Eq.~\eqref{Htot} starting with $H_0$ given by a RM model, Eq.~\eqref{RMvec}, is typically not a nearest-neighbor model which is obvious from the generically $k$-dependent $R^2(k,t)$ denominator of the CD term in~\eqref{Htot}. Since $H_{CD}(k,t)$ is generically a smooth function of $k$, by elementary Fourier analysis, it is exponentially localized in real space; however, the localization length may be hundreds of sites long. We construct a general class of nearest neighbor CD protocols, by demanding that the total Hamiltonian $H$ have a Bloch vector $\bm{u}(k,t)$ of the form
\begin{align}\label{ucons}
  \bm{u}(k,t)=
  \scalebox{0.95}{$\displaystyle
  \begin{bmatrix}
    -J_1(t)-J_2(t)\cos(2ka) - \delta J_3(t)\sin(2ka) \\
    -\delta J_1(t)-J_2(t)\sin(2ka)-\delta J_4(t)\cos(2ka) \\
    -\Delta(t)
  \end{bmatrix}$}
\end{align}

This construction results in a protocol that is nearest-neighbor (NN) in the sense that it only pairwise couples adjacent cells.  To obtain site-pairwise-NN, as in the RM model, we require the additional condition $\delta J_3(t) = \delta J_4(t) \equiv \delta J_2(t)$. Throughout the text NN means site-NN. We then solve for $\bm{R}$ using Eq.~\eqref{Htot} to determine $H_0(k,t)$. We demand that $\bm{R}$ begins and ends as a given, arbitrary, RM model. However, at intermediate times, $\bm{R}$ will not generically be a RM model.

We require that the system start in the ground state of the Hamiltonian $H_0(k,t\leq 0)$ and return to the ground state at the end of the protocol, which leads to the constraint
\begin{equation}\label{eqbc}
    \bm{R}(k,0)=\bm{u}(k,t\leq 0)=\bm{u}(k,t \geq T)=\bm{R}(k,T).
\end{equation}
This condition is critical to ensure that $\bm{R}$ is periodic in $k$ and $t$ during the drive, which leads to quantization of the pumped charge. While the first half of this condition is simply set as an initial condition for $\bm{R}$ in Eq.~\ref{Htot} viewed as a time-evolution equation for $\bm{R}(k,t)$, the second half constrains the final value $\bm{R}(k,t)$ of the solution. We will construct a solution that satisfies both Eqs.~\eqref{ucons} and~\eqref{eqbc}.

To determine the appropriate $\bm{R}(k,t)$, we start by noting that the equation for $H(k,t)$ (i.e. Eq.~\eqref{Htot}) is explicitly written as a differential equation  $\bm{u} = \left(\bm{R}+\frac{\hbar}{2}\hat{\bm{R}}\times\partial_t\hat{\bm{R}}\right)$. This equation can be rewritten as
\begin{align}\label{veceom}
    \partial_t\hat{\bm{R}}=\frac{2}{\hbar}\bm{u}\times\hat{\bm{R}} 
\end{align}
with the magnitude $R = |\bm{R}|$ determined by 
\begin{equation}\label{vecmag}
    R=\bm{u}\cdot\hat{\bm{R}}. 
\end{equation}

We develop a general method to solve Eqs.~\eqref{veceom} and~\eqref{vecmag} subject to Eqs.~\eqref{eqbc} and~\eqref{ucons} with an initial condition of a generic Rice-Mele model Eq.~\eqref{RMvec} (whose bands are not flat), that, moreover, produces non-zero, quantized pumping.  We use numerical optimization to construct $\bm{u}(k,t)$ of the nearest-neighbor form Eq.~\eqref{ucons}, where the solution $\bm{R}(k,t)$ interpolates between the initial Rice-Mele model $\bm{u}(k,t<0) = \bm{R}(k,0)$ and the starting point of our bucket-brigade protocol at $t=T/4$: $\hat{\bm{R}}(k,T/4) = (0,0,-1)$. We then evolve the system over $[\,T/4,3T/4)$ using the analytic CD-bucket-brigade protocol, and complete the cycle by evolving the system under the time-reversed $\tilde{\bm{u}}(k,t) = \bm{u}(k,T-t)$. The details of the optimization are given in Appendix~\ref{appendix:CD_protocol}.

\begin{figure*}[h!t]
    \centering

    \begin{subfigure}{\textwidth}
        \centering
        \includegraphics[width=1.0\textwidth]{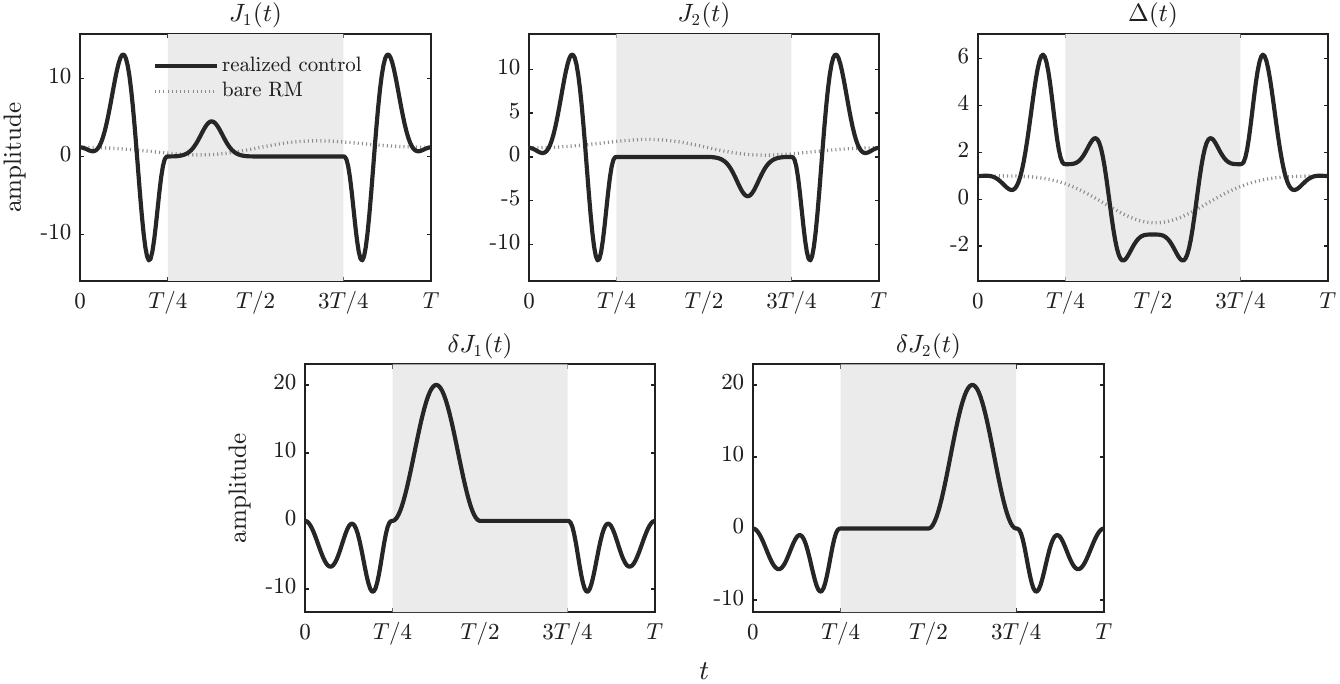}
        \caption{}
        \label{fig:control_functions}
    \end{subfigure}

    \vspace{0.5em} 

    \begin{subfigure}{\textwidth}
        \centering
        \includegraphics[width=1.0\textwidth]{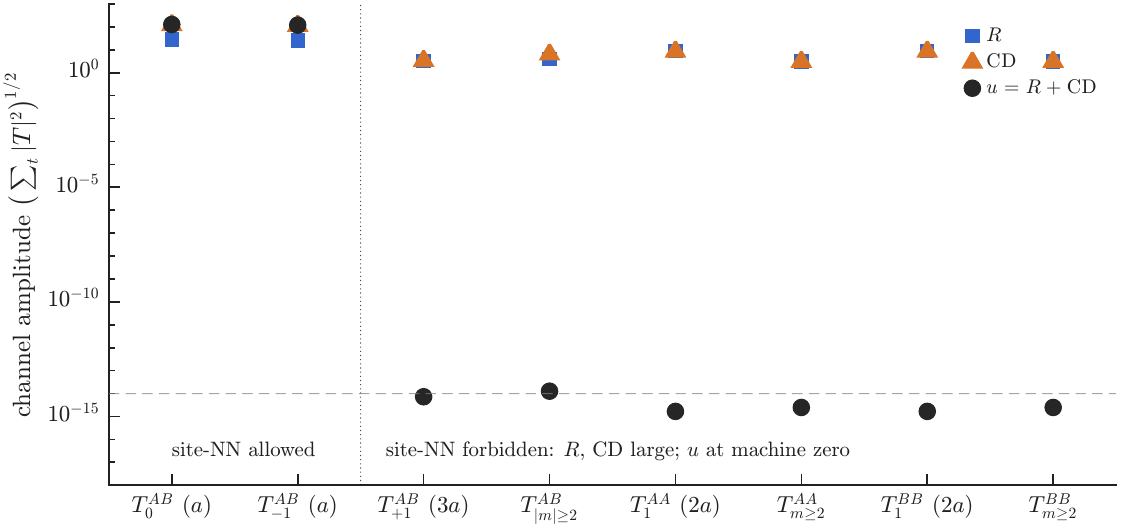}
        \caption{}
        \label{fig:siteNN_cancellation}
    \end{subfigure}

    \caption{\textbf{(a)} Control functions which are the components of $\bm{u}$, which begin and end in an arbitrary RM model (dashed line ``bare RM"). The gray region corresponds the the BBCD protocol.  \textbf{(b)} The real-space hopping channels of $\bm{R}$, the CD term $(1/2)\hat{\bm{R}}\times\partial_t\hat{\bm{R}}$ the total Bloch vector $\bm{u}$. We see that former involve long range hopping, but exactly cancel to produce a site-NN control protocol $\bm{u}$. Each $T^{XY}_m$ is the coupling between a site on sublattice $X$ and a site on $Y$ that are $m$ unit cells apart. In particular, $T^{AB}_m(t) = \langle H_{AB}(k,t)e^{-2iamk}\rangle = (1/N_k)\sum_{k}(R_x(k,t)-iR_y(k,t))$, and $T^{AA}_m(t) = \langle H_{AA}(k,t)e^{-2iamk}\rangle = (1/N_k)\sum_{k}(R_z(k,t)) = -T^{BB}_m(t)$ which are the real space hopping strengths as functions of time. The log-scale y-axis is the sum over a cycle of the squares of these hopping strengths. Plots produced for $\omega = 10$, which exceeds the adiabatic limit set by $0.47 < min_k|R|< 8.1$}
    \label{fig:ControlWeight}

\end{figure*}

The optimization constructs a vector $\bm{R}(k,t)$ that cancels the non-nearest neighbor contributions of its corresponding CD term, as shown in Fig.~\ref{fig:siteNN_cancellation}, as well as the protocol $\bm{u}(k,t)$ that interpolates between the initial ground state of an arbitrary RM model to the initial ground state of the BBCD protocol: an analytic, simple, nearest-neighbor $\bm{R}(k,t)$ with flat bands that lead to a corresponding site-NN CD term, see Fig.~\ref{fig:control_functions}. The result is an overall protocol that given an arbitrary RM initial ground state, produces non-trivial, finite-time quantized pumping, with a site-NN total Hamiltonian, $\bm{u}(k,t)\cdot\bm{\sigma}$, which we again underscore is the actual Hamiltonian that would be implemented experimentally. The vector $\bm{R}$, and the corresponding $H_0$, are thus computed quantities used to determine the Chern number, the adiabatic limit for the dynamics, and pumped charge quantization.

\begin{figure*}[h!t]
\includegraphics[width=0.8\textwidth]{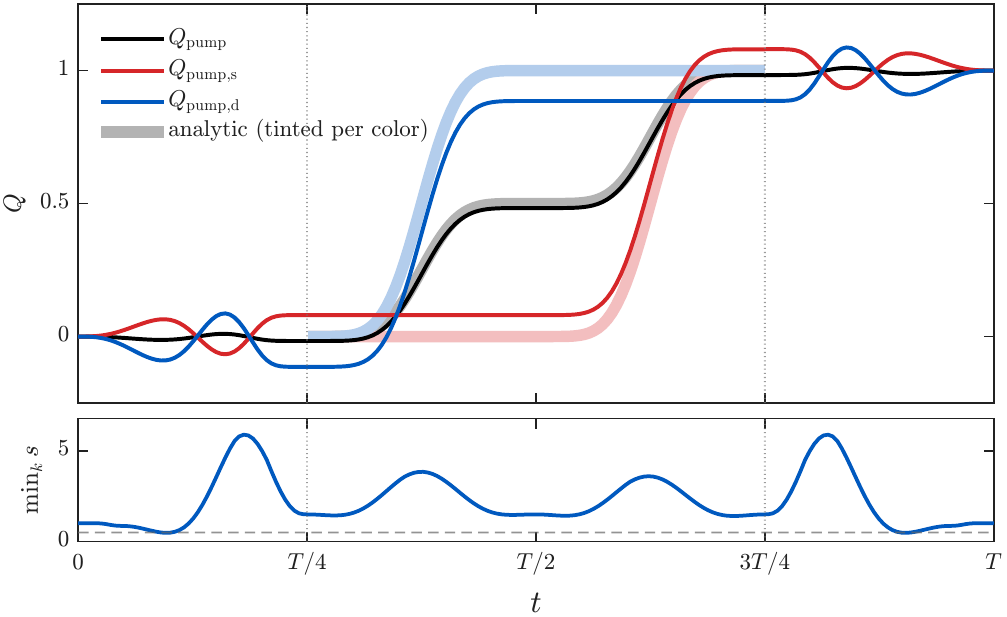}
\caption{$Q_{pump}(t)$ of Eq.~\eqref{eqQpumpC} and $Q_{pump,b}(t)$ of Eq.~\eqref{Qcontrol} for $\omega =$ $10$, an order of magnitude larger than the adiabatic limit which is determined by the gap of $s \equiv|R(k,t)| = \bm{u}\cdot\hat{\bm{R}}$. The pumped charge produced by the optimized protocol overlaid with the BBCD protocol during $T/4$ to $3T/4$. Note the pumped charged produced by the optimization protocol is quantized over a cycle $T$. The pumped charge departs from the analytic BBCD results (Eqs.~\eqref{Qavgcontrol} and~\eqref{Qcontrol}) between $T/4$ and $3T/4$ due to imperfect initialization of the BBCD protocol from the optimizer. Physically, the particle at site $A$ is sequentially pumped to site $B$ in the same unit cell and then pumped to site $A$ in the next unit cell. As the optimization sends both $J_1$ and $J_2$ of a generic RM model to approximately $0$ to reach the initial step of the control freak, slight forward/backward pumping is produced in the first and final legs from due to the optimization protocol and its time-reversal}
\label{Q_pump}
\end{figure*}

We emphasize that the adiabatic regime of the system is not set by the total Hamiltonian; thus, $|\bm{u}|$ does not change the adiabatic time scale of the system's dynamics. We discuss this further in Appendix~\ref{appendix:H0vsCD}. The gap of $H_0(t)$ is $2|R(k,t)|$. At $t=0$ this is the gap of the initial RM model, and at $t = T/4$ this is the gap of the bucket-brigade protocol: $|R(T/4)| = e^{\lambda(T/4)}$. The bottom panel of Figure~\ref{Q_pump} shows the minimum value of the half of the band gap $\Delta E = 2|R|$, which determines the adiabatic limit: $2\pi/T_{ad} = \omega_{ad} \ll \Delta E/\hbar$ which we see ranges from $0.47$ to $8.1$ in the plot. Thus, $\omega = 10 > \Delta E/\hbar \gg \omega_{ad}$ unambiguously exceeds the adiabatic regime.

Figure~\ref{Q_pump} displays the resulting pumped charge produced by the protocol found via optimization. Plots are produced for $\omega = 10$. The slight departure from the pumped charge equaling one half at one half of the period is to be expected due to the imperfect initialization produced by the optimization step: the system is not exactly in the ground state of bucket brigade RM at the end of the optimization at $T/4$. This is why the particle transport is not perfectly quantized at the end of the bucket brigade steps at $3T/4$. However, Figure~\ref{Q_pump} shows that the pumped charge is manifestly quantized at the end of the entire drive period, which includes four steps: optimization, the two bucket brigade steps, and the time-reversed optimization protocol all involving only nearest neighbor couplings.

\section{Discussion}\label{discussion}Starting from a Rice-Mele model, we have derived an exact $H_{CD}$ protocol and  have shown analytically that it preserves the charge pumped across each bond as a Chern number. We did this by integrating the current operator to find the pumped charge $Q_{pump,b}(t)$ for the two bond types, $s$ and $d$, in the lattice shown in Fig.~\ref{fig:RM}. However, this exact $H_{CD}$ generally involves higher harmonics of $k$, which in real space correspond to long-range hopping terms. Such terms are challenging to actualize experimentally. The total Hamiltonian $H = H_0 + H_{CD}$ is what is actually implemented in an experiment.
We thus considered the possibility of restricting $H(t)$ to only include nearest-neighbor hopping terms through Eq.~\eqref{ucons}.

Within this nearest-neighbor constraint we found an adiabatic bucket brigade protocol of the RM model which produced an analytically soluble example of fast, non-trivial, pumping as manifest in Eq.~\eqref{Qavgcontrol} and~\eqref{Qcontrol}. We then showed that BBCD protocol reproduces the topological robustness of adiabatic Thouless pumps to disorder, away from the adiabatic limit. We moreover showed that the BBCD protocol possess a CD-robustness: for sufficiently rapid driving, it is remarkably resilient to noise and onsite disorder. 
We can understand this resilience via the infinitely fast driving limit: the very large CD term acts over an infinitesimal time which dominates the dynamics, while the effects of the finite $H_0$ and $H_{\mathrm{dis}}$ are negligible. Our numerics demonstrate that this limit is effectively achieved when the drive speed  is one order of magnitude larger than the time-scale set by strength of the noise and disorder.

We then developed a general numerical optimization method to construct a nearest-neighbor, topologically non-trivial, $H_{CD}$ protocol that rapidly drives a system initialized in the ground state of an arbitrary clean Rice-Mele model around a cycle, while maintaining the system in its ground state at all times, leading to fast, non-trivial, quantized charge pumping  thus demonstrating the generality of constructing nearest neighbor CD protocols. Both the exact bucket brigade and the full optimization protocols possess a clear physical interpretation of the time-dependence of the charge pumped across the bonds: overall, the charge is first pumped from site $A$ to $B$ within a unit cell and then to site $A$ in the next unit cell.

 The single-cycle fast Thouless pump developed here in Eq.~\eqref{Htot}, for any non-topologically trivial Rice-Mele $H_0$ can also be described as a special case of a Floquet system (via executing many pump cycles) with a non-trivial topological number \cite{kitagawa2010topological}. Since Floquet systems are characterized by quasi-energies, these systems do not come with a natural ground state that is required for a quasi-adiabatic Thouless pump. The result from Eq.~\eqref{Htot} avoids this challenge, since the initial state under the drive is the ground state of $H_0$. Our fast Thouless pump protocols have the property that the Floquet state
matches the instantaneous ground state of $H_0$ for at least one point in time in the cycle interval. The protocol's non-trivial Floquet topological invariant is another argument for why the protocol provides a quantized Thouless pump over a single pumping cycle as
opposed to a periodic drive. We note a general point: we can readily find Floquet eigenstates using CD methods in periodic systems. In Ref.~\cite{malikis2022ideal}, Malikis and Cheianov (MC) solve the Floquet initialization problem for a constrained class of RM models characterized by specific parameter-space trajectories for the hopping coefficients $J_1(t)$, $J_2(t)$ and onsite potential $\Delta(t)$. MC define $\delta(t) = |J_1(t)| - |J_2(t)|$, and their framework implies that if $\partial_t \Delta( t ) = -n\lambda\delta( t ) \sqrt{1+4\delta( t )^2  }$ then the RM Hamiltonian can be mapped to the Sinh-Gordon equation, where $n\in \mathbb{Z}$ and $\lambda$ is the fundamental period of a solution to the Sinh-Gordon equation. The standard analytic RM protocol, which is the initial condition of our optimization, does not satisfy this constraint, nor does the bucket brigade protocol.

We note a key difference between geometric~\cite{citro2023thouless} and topological pumps: in a topological pump, the system's two parameters are automatically uniformly sampled for any pump protocol that preserves the system as an insulator. The many-body character of these systems, which is manifest in the uniform filling of the bands is essential for the corresponding uniform integration over $k$ (in the clean system), which maps the pumped charge onto to a Chern number and thus is fundamental to producing the exact, topological, quantization of pumped charge. By contrast, for geometric pumps, which are inherently single-particle systems, the uniform sampling can be affected by time-variations of parameters; the resulting expression for pumped charge is then given by:
\begin{equation}
\begin{split}
Q_{pump} &= \iint dk dt f(k,t) B(k,t) 
\\
&= C + \iint dk dt (f(k,t)-1) B(k,t)
\end{split}
\end{equation}
where $B(k,t)$ is Berry curvature, $Q_{pump}$ is pumped charge, $f(k,t)$ is a sampling function, and $C$ is the first Chern number.
The final term is the so-called Berry curvature dipole. Thus, the pumped charge manifestly depends on geometric quantities. For example, Cooper pair~\cite{erdman2019fast} and BEC~\cite{lu2016geometrical} pumps operate over many realizations of single-particle $0$d, open systems, where there is not an automatic, physical enforcement of uniform sampling over both periodic parameters of these systems. Only one is physically justified: the time-integral to extract displacement from velocity. To uniformly sample over the second parameter, and thus map onto a topological invariant, requires either an added lattice tilt, i.e. an external force~\cite{ke2020topological}, or an ensemble average~\cite{erdman2019fast} of a gauge invariant quantity, which both the Cooper pair and BEC systems inherently lack without additional assumptions. Thus, these pumps are not robust to the choice of protocol.

The clean RM model's translational invariance allows for the standard diagonalization procedure in $k$-space as in Eq.~\eqref{RMHfull}, and Eq.~\eqref{Htotk}, within which the calculations are those of single particle physics and the many-body character of the system comes from the filling. The resulting $H_{CD}$ is smooth in $k$ and therefore generically exponentially localized in real space. The exponential localization of lattice $H_{CD}$ operators is consistent with previous literature, however, to the best our knowledge, has not been previously noted.
The work of~\cite{sels2017minimizing} considers a $1$d lattice model of free fermions, and argues that the generic exact CD term is non-local, in that it involves long range hopping. They develop an approximate CD operator which is NN local by construction. The RM model also includes time varying hopping in addition to the time-varying onsite potentials they consider. Our method to construct an exact, NN CD protocol for a free fermion lattice system appears to be generalizable to any lattice with time-dependent couplings. While bucket brigade methods are well established in non-adiabatic many-body dynamics, we emphasize the value of their application to STA methods to produce an exact, nearest-neighbor free fermion protocol. We can define a class of adiabatic protocols akin to bucket brigade, that alternate between a fully disconnected lattice and steps involving the partially connected lattice. We conjecture that the corresponding CD protocols will also be highly localized in real space.
Given that essentially all experimental implementations of Rice-Mele Thouless pumps to date involve only nearest-neighbor interactions, these fast, robust, nearest-neighbor, quantized pumping protocols may provide near-term opportunities for experimental realizations of finite-time topologically quantized pumps. Since the BBCD protocol operates in the dimerized limit, it corresponds to executing finite-time bit-flip cycles along the RM chain. Moreover, the BBCD drive possesses an inherent measure of errors to these bit-flip cycles: the quantization of pumped charge, which may prove useful for quantum computational hardware.

\textbf{Note added}: In the process of completing the preprint of this manuscript~\cite{chiel2024fast}, we became aware of the preprint by Liu {\it et al}~\cite{liu2024shortcuts} which has since been published~\cite{liu2025shortcuts} during the review of this manuscript, who also apply counterdiabatic methods to the Rice-Mele model. 
Their work marshals the Wannier center formalism to numerically demonstrate pumped charge quantization obtained by the counterdiabatic field $\bm{u}(k,t)$, which matches our Eq.~\eqref{Htot}, and complements our analytic results which prove quantization of charge pumping via integrating the current operator. Wherever our results overlap, they agree. They also give numerical evidence for robustness of their long-range hopping protocol to unknown lattice-translationally invariant noise in their system's parameters. However, there are number of key additional results we present. Eq.~\eqref{Htot} generically requires long-range hopping, and so we constructed an exact NN CD field, and in addition show analytically and numerically the robustness of the resulting non-adiabatic pumped charge to known weak, onsite disorder that breaks lattice-translation invariance, as in Thouless' original construction. In addition, we analytically justify and numerically demonstrate the regimes of the CD drive's robustness to unknown onsite disorder and noise in the model's parameters that break lattice-translation invariance, in achieving both rapid state transfer and quantized pumped charge. We also address the impact of thermal fluctuations. Lastly, we construct NN CD fields for any initial RM model, via numerical optimization.

\textbf{Acknowledgments}: JC is supported by the National Science Foundation Graduate Research Fellowship Program under Grant No. 2236417, the National Science Foundation under Grant No. 2127900, and the John Templeton Foundation under Award No. 63626. CJ acknowledges support by the U.S. National Science Foundation under Grant No. 2127900.
Any opinions, findings, and conclusions or recommendations expressed in this material
are those of the authors and do not necessarily reflect the views of the National Science
Foundation.
J.S. acknowledges support from the Joint Quantum Institute
and support by the Laboratory for Physical Sciences through its continuous support of the Condensed Matter Theory Center at the University of Maryland.
JC gratefully acknowledges Michael Hinczewski, Harsh Mathur, and Sebastian Deffner for helpful conversations. 

	\bibliography{bbl} 

\clearpage
\appendix

\section{Current operator}\label{appendix:current}

Consider an arbitrary non-local lattice Hamiltonian defined over a set of points, corresponding to uniformly spaced, with spacing $a$, atomic sites in $1$d of length $L$, with periodic boundary conditions: 
\begin{equation}\label{nonlocH}
i\hbar\partial_t\Psi(x,t) = \sum_{x'} H(x,x',t)\Psi(x',t) \, .
\end{equation}
where $x,x' \in [-\frac{L}{2},\frac{L}{2})$. We minimally couple our system to a vector potential $A(y)$ by writing
\begin{equation}\label{Hnonloc1}
H(x,x',t)[A] = H(x,x',t) \, e^{-i \frac{e}{\hbar}\oint_{x'}^{x} dyA(y)},
\end{equation}
where $H(x,x',t)$ is given by Eq.~\eqref{nonlocH}, $y\in(-\frac{L}{2},\frac{L}{2})$ is a continuous position defined on the bonds between atomic sites, and $e = |e|$ is the electron charge's magnitude. We define the ``shortest path" line integral on the circle, $\oint_{x'}^{x}dy \equiv \int_{\tilde{x}'}^{\tilde{x}}dy$ where the points $x,x'$ are mapped to $\tilde{x},\tilde{x}'$ which are equivalent modulo $L$, such that $|\tilde{x}-\tilde{x}'| < L/2$. We note that $\hat{H}[A]$ is Hermitian, since $H(x,x',t)[A] = (H(x',x,t))^{*}$.

The current operator is given by the functional derivative~\cite{nourafkan2018hall},
\begin{equation}
    \bra{x}\hat{J}(y)\ket{x'} = \left.\frac{\delta H(x,x',t)[A]}{\delta A(y)}\right|_{A\to 0} 
\end{equation}

Taking the functional derivative we find

\begin{equation}\label{eq:Jreal}
\scalebox{0.93}{$\displaystyle
\begin{split}
\bra{x}\hat{J}(y)\ket{x'}
&= -i\frac{e}{\hbar}H(x,x',t)
\\
&\Biggl(\Theta(f(y-x'))\Theta\bigl(f(x-x')-f(y-x')\bigr)
\\
&\quad-\Theta(f(y-x))\Theta\bigl(f(x'-x)-f(y-x)\bigr)\Biggr).
\end{split}$}
\end{equation}

Here, $\Theta(x)$ is the Heaviside step function and $f(x) = x \text{ for } x \in [-\frac{L}{2},\frac{L}{2})$, which is monotonic on this interval, and $f(x+L) = f(x)$. We note that we get a $+1$ contribution to the current for $x,x',y$ that satisfy the following relation: $0 < f(y-x') < f(x-x')$ for $0 < x - x' < \frac{L}{2}$ (i.e. clockwise flow from $x'$ to $x$) and a $-1$ contribution for $0<f(y-x)\leq f(x'-x)$ for $0 < x' - x \leq \frac{L}{2}$ (counterclockwise flow from $x'$ to $x$).

It suffices to consider $|x - x'| < \frac{L}{2}$ since the periodicity of $f$ determines $|x - x'| > \frac{L}{2}$ from this case.
 Using $\Theta(-x) = 1-\Theta(x)$ and the definition of $f$ we can write
\begin{equation}
\begin{split}
&\bra{x}\hat{J}(y)\ket{x'} = -i\frac{e}{\hbar}H(x,x',t)
\\
&\Biggl(\Theta(y-x')\Theta(x-y)-\Theta(y-x)\Theta(x'-y)\Biggr)
\\
&=-i\frac{e}{\hbar}H(x,x',t)\Biggl(\Theta(y-x')(1-\Theta(y-x))
\\
&-\Theta(y-x)(1-\Theta(y-x'))\Biggr)
\\
&=-i\frac{e}{\hbar}\Bigl(\Theta(y-x')-\Theta(y-x)\Bigr)H(x,x',t).
\end{split}
\end{equation}

We show that Eq.~\eqref{eq:Jreal} agrees with the notion of current as the flow of charge. 
The center points of the two bonds that meet at site $x$ are $x_\pm \equiv x \pm a/2$. Given sites $x, x_1, x_2$ with $ |x_2-x_1|< \frac{L}{2}$, we have
\begin{equation}\label{eq:diff}
\begin{split}
    \bra{x_2}\hat J(x_-)\ket{x_1} - \bra{x_2}\hat J(x_+)\ket{x_1} \\
    = -i \frac{e}{\hbar} \left( \delta_{x,x_2} - \delta_{x,x_1} \right) H(x_2,x_1,t).
\end{split}
\end{equation}
We then have:
\begin{equation}\label{eq:continuity}
\begin{split}
   &\langle \hat J(x_-) \rangle - \langle \hat J(x_+) \rangle \\
    &= \sum_{x_1,x_2} \Psi^*(x_2,t) \bra{x_2} \Bigl[\hat J(x_-) - \hat J(x_+)\Bigr] \ket{x_1} \Psi(x_1,t) \\
    &= -i\frac{e}{\hbar} \Bigl[ \sum_{x_1} \Psi^*(x,t) H(x,x_1,t) \Psi(x_1,t) \\
    & \qquad - \sum_{x_2} \Psi^*(x_2,t) H(x_2,x,t) \Psi(x,t) \Bigr] \\
    &= e\, [\Psi^*(x,t) \,\partial_t \Psi(x,t) + \left(\partial_t \Psi^*(x,t)\right)\Psi(x,t)] \\
    &= e\, \partial_t \vert \Psi(x,t) \vert^2 = \partial_t Q(x,t)
\end{split}
\end{equation}
using $H(x,x_1,t) = (H(x_1,x,t))^*$ and Eqs.~\eqref{nonlocH}, \eqref{eq:diff} and where $Q(x,t) = e|\Psi(x,t)|^2$ .
Eq.~\eqref{eq:continuity} has a simple interpretation: charge accumulates at site $x$ at a rate given by the difference between the inflow and outflow of current. 

For any function $g$ that is defined over a bond when $x$ is on a site, we can define a discrete derivative $(\delta_x g)(x)=g(x+a/2)-g(x-a/2)$. We can thus rewrite Eq.~\eqref{eq:continuity} as a discrete version of the continuity equation:
\begin{align}
-\delta_x\langle\hat{J}\rangle = \, \partial_t Q(x,t).
\end{align}

For arbitrary $x$ and $x'$, $\bra{x}\hat{J}(y)\ket{x'}$ is manifestly independent of $y$ within any bond between two neighboring sites. Therefore, it suffices to consider this operator over a discrete grid of $y$ values -- one value between each pair of adjacent sites. For the lattice shown in Fig.~\ref{fig:RM}, we choose the intracell bonds (d, double green lines) at positions $y_{j,d} = a(2j+1/2)$ and intercell bonds (s, single green lines) at positions $y_{j,s} = a(2j-1/2)$. We denote the set of all of the bonds as $Y$. Thus, the sites and bonds are located at integer and half-integers multiples of $a$, respectively. The current along the bond that contains $y\in Y$ is ${\rm Tr}\left[\hat{J}(y)\hat{\rho}\right]$, where $\hat{\rho}$ is a density operator describing the system's state.
Averaging the current over all of the system's bonds, we obtain
\begin{equation}\label{eq:javggen}
\begin{split}
j_{avg}(t)&=\frac{a}{L}\sum_{y \in Y}{\rm Tr}\left[\hat{J}(y)\hat{\rho}\right] 
\\
&=\frac{a}{L}\sum_{y \in Y}\sum_{x,x'}\bra{x}\hat{J}(y)\ket{x'}\rho(x',x,t)
\\
&=-\frac{a}{L}\frac{ie}{\hbar}\sum_{x,x'}f\left(\frac{x-x'}{a}\right)
\\
&H(x,x',t)\rho(x',x,t)
\end{split}
\end{equation}
using the following result:
\begin{equation}
\begin{split}
f\left(\frac{x-x'}{a}\right)&= 
\sum_{y\in Y}\Biggl(\Theta(f(y-x'))
\\
&\Theta\biggl(f(x-x')-f(y-x')\biggr)
\\
&-\Theta(f(y-x))
\\
&\Theta
\biggl(f(x'-x)-f(y-x)\biggr)\Biggr).
\end{split}
\end{equation}
We now specialize to a translationally invariant lattice with basis: a repeating lattice with $\mu=2$ atoms per cell, so that $L=a\mu N$.
We take $x = r + c$ and $x' = r' + c'$ to represent the positions of atoms, as, for example, in Fig.~\ref{fig:RM}, where $r,r' \in \{ 2aj \,|\, j\in \left[ -N/2 , N/2 \right)\cap \mathbb{Z}\}$ and $c,c' \in \{0,a\}$.
Defining $\bar{r} \equiv r - r'$ and using the translational invariance of $H$, our system's density matrix, as well as $f(\frac{x-x'}{a})$, we have 
\begin{equation}\label{avgcurrent}
    j_{avg}(t) = \frac{i e}{\hbar a \mu}\sum_{\bar{r},c,c'}(\bar{r}+c-c')H(\bar{r}+c,c',t)\rho(c',\bar{r}+c,t)
\end{equation}
Using Fourier series, we now recast the real space Hamiltonian in terms of the Bloch Hamiltonian as $H(\bar{r}+c,c',t)= \frac{1}{N}\sum_k e^{-i k \bar{r}}\tilde{H}_{k}(c,c',t)$, where $\tilde{H}_k(c,c',t) =\sum_{\bar{r}}e^{i k \bar{r}}H(\bar{r}+c,c',t)$, and $\rho(\bar{r}+c,c',t) = \frac{1}{N}\sum_k e^{-i k \bar{r}}\tilde{\rho}_{k}(c,c',t)$. 

Plugging these expressions into Eq.~\eqref{avgcurrent}, we find
\begin{equation}\label{current_calc}
\begin{split}
j_{avg}(t)=&\frac{ie}{\hbar N^2}\sum_{k,k'}
\sum_{\bar{r},c,c'}\biggl((\bar{r}+c-c')e^{-i k \bar{r}}\tilde{H}_{k}(c,c',t)\\
&e^{i k' \bar{r}}
\tilde{\rho}_{k'}^*(c,c',t)\biggr)\\
&=\frac{ie}{\hbar N^2}\sum_{k,k'}\sum_{\bar{r},c,c'}\biggl(\tilde{H}_{k}(c,c',t)\tilde{\rho}^*_{k'}(c,c',t)
\\
&(\bar{r}+c-c')e^{-i(k-k') \bar{r}}\biggr)
\\
&=\frac{ie}{\hbar N^2}\sum_{k,k'}
\sum_{\bar{r}} \biggl(\bar{r} {\rm Tr}\left(\hat{\tilde{H}}_{k}(t)\hat{\tilde{\rho}}^{\dagger}_{k'}(t)\right)
\\
&+{\rm Tr}\left([\hat{c},\hat{\tilde{H}}_{k}(t)]\hat{\tilde{\rho}}^{\dagger}_{k'}(t)\right)\biggr)e^{-i( k-k') \bar{r}}
\end{split}
\end{equation}

Here, $\hat{c}$ is the position operator projected onto the basis sites within the unit cell. We can further process the second term in the final line using the identity 
\begin{equation}\label{delta}
\sum_{\bar{r}}e^{-i( k-k') \bar{r}} = N\delta_{k,k'},
\end{equation}
the cyclic property of the trace, and the von Neumann equation, to find $-\frac{e}{ N}\sum_k\partial_t\left({\rm Tr}\left(\hat{c}\hat{\tilde{\rho}}^{\dagger}_{k}(t)\right)\right)$.

For the first term, we here assume that $\tilde{H}_k$ is a differentiable function of $k$ and thus the first term has a well defined first partial derivative w.r.t. $k$. The sum over $k$ then adds together the evaluations of this $k$-continuous function along a discrete mesh of $k \in \{k_i\}_i$ values and hence we can rewrite the first term as
\begin{equation}
    \begin{split}
        &{\scriptstyle\frac{ie}{\hbar N^2}\sum_{i,j,\bar{r}}\left( 
    {\rm Tr}\left(\hat{\tilde{H}}_{k}(t)\hat{\tilde{\rho}}^{\dagger}_{k'}(t)\right)i\partial_k\right)
    e^{-i( k-k') \bar{r}}\Biggr|_{k=k_i, k' = k_j}}
\\
&{\scriptstyle=\frac{e}{\hbar N^2}\biggl(\sum_{i,j} 
{\rm Tr}\left((\partial_k \hat{\tilde{H}}_{k}(t))\hat{\tilde{\rho}}^{\dagger}_{k'}(t)\right)\sum_{\bar{r}}e^{-i( k-k') \bar{r}}}
\\
&{\scriptstyle-\frac{e}{\hbar N^2}\sum_{i,j} \tilde{\rho}^{\dagger}_{k'}(t)\sum_{\bar{r}}\partial_k\left(\hat{\tilde{H}}_k(t) e^{-i( k-k') \bar{r}}\right)\biggr)\Biggr|_{k=k_i,k'=k_j}}
    \end{split}
\end{equation}

where the equality follows from the product rule, 
\begin{equation}
\begin{split}
\hat{\tilde{H}}_k(t)\partial_k e^{-i( k-k') \bar{r}} &= -e^{-i( k-k') \bar{r}}\partial_k\hat{\tilde{H}}_k(t) 
\\&+\partial_k\left(\hat{\tilde{H}}_k(t) e^{-i( k-k') \bar{r}}\right). 
\end{split}
\end{equation}
Using~\eqref{delta}, the first term becomes $\frac{e}{\hbar N}\sum_{i} 
{\rm Tr}\left((\partial_k \hat{\tilde{H}}_{k}(t))\hat{\tilde{\rho}}^{\dagger}_{k}(t)\right)\biggr|_{k=k_i}$. 

We can approximate the total derivative of the second term as a finite difference, and rewriting the sum over $i$ as a Riemann sum results in a telescoping sum which gives $0$, by $k$-periodicity, up to an error term that vanishes as the $k$-mesh size goes to $0$, i.e. in the thermodynamic limit, $N\to \infty$.

Substituting the Bloch Hamiltonian $\hat{H}(k,t)$ of Eq.~\eqref{Htot} for the unit cell Hamiltonian $\tilde{H}_k(c,c',t)$ along with with $\hat{\rho}_k(t) = \frac{1}{2}\left(\mathbb{I} - \hat{\bm{R}}(k,t)\cdot\bm{\sigma}\right)$ as the single particle ground-state density matrix, and taking the thermodynamic limit, Eq.~\eqref{current_calc} simplifies to:
\begin{equation}
\begin{split}\label{unit_cell_cur1}
    j_{avg}(t)= \frac{e}{\hbar}\int \frac{dk}{2\pi}\left(\partial_k\bm{u}\cdot\hat{\bm{R}}-\hbar\partial_t{\rm Tr}\left[\hat{c}\hat{\rho}_k^{\dagger}\right]\right),
    \end{split}
\end{equation}
where $\hat{\bm{R}} = \bm{R}/R$. 
We note that in the quasi-adiabatic limit our results match the standard form for average current~\cite{souza2004dynamics}.

\section{Numerically optimized CD nearest-neighbor protocol}\label{appendix:CD_protocol}
Using the BBCD protocol, we develop a general method to solve Eqs.~\eqref{veceom} and~\eqref{vecmag} subject to Eqs.~\eqref{eqbc} and~\eqref{ucons} with an initial condition of a generic Rice-Mele model Eq.~\eqref{RMvec}. We use numerical optimization to construct $\bm{u}(k,t)$ of the nearest-neighbor form Eq.~\eqref{ucons}, where the solution $\bm{R}(k,t)$ interpolates between the initial Rice-Mele model $\bm{u}(k,t<0) = \bm{R}(k,0)$ and the starting point of our control freak protocol at $t=T/4$: $\hat{\bm{R}}(k,T/4) = (0,0,-1)$, where we now simply shift the continuity conditions on $\theta(t)$ and $\dot{\theta}(t)$ to correspond to $T/4, \, T/2,\, 3T/4$. We then evolve the system over $[\,T/4,3T/4)$ using the BBCD protocol, and complete the cycle by evolving the system under the time-reversed $\tilde{\bm{u}}(k,t) = \bm{u}(k,T-t)$. For $\bm{u}(k,t)$ to be nearest-neighbor in real space, it must be a smooth function of $k$.
To optimize, we vary the five nearest-neighbor control functions
$g_i \in \{J_1, J_2, \Delta, \delta J_2,\delta J_1\}$, writing
\begin{equation}\label{eq:ansatz}
  g_i(\varphi(t)) = g^{\rm RM}_i(\varphi(t)) + \sigma_i\, d_i\!\big(\varphi(t)\big),
\end{equation}
where 
\begin{equation}
    d_i(\varphi) = \sum_{n=1}^{N_c}\Big[B^{s}_{i,n}\sin n\varphi + B^{c}_{i,n}\big(\cos n\varphi - 1\big)\Big]
\end{equation}
and $\sigma=(1,-1,1,1,1)$ and $\varphi(t)=\pi\!\left(1-\cos\tfrac{\omega t}{2}\right)$, with $N_c$ the number of $\varphi$-harmonics.
The $g_i$ control functions enter the protocol via
\begin{align}\label{eq:uNN}
  u_x(k,t) &= -J_1(t) - J_2(t)\cos 2ak - \delta J_2(t) \sin 2ak, \nonumber\\
  u_y(k,t) &= -\delta J_1(t)-J_2(t)\sin 2ak + \delta J_2(t)\cos 2ak 
  \nonumber\\
  u_z(k,t)  &= -\Delta(t) .
\end{align}
The reference functions $g^{RM} = \{J_1^{RM}(t),J^{RM}_2(t),\Delta^{RM}(t),0,0\}$ are the bare Rice--Mele drive, with
$J^{(0)}_{1/2}(t) = J_0 \pm \delta_0\cos(\varphi(t)+\phi_{\rm shift})$,
$\Delta^{(0)}(t) = \Delta_0\sin(\varphi(t)+\phi_{\rm shift})$,
$\delta J^{(0)}_{3,4}=0$. The time shift $\phi_{shift}$ allows for general initial starting points. 
The $\big(\cos n\varphi - 1\big)$ basis vanishes at $t=0$, so the system
is initialized exactly in the Rice--Mele model and since $\dot\varphi(0)=0$, we also have
$\partial_t\bm{R}(k,t)|_{t=0}=0$.  We only optimize over $t\in[0,T/4]$ (which is justified for sufficiently small optimization errors). On $[T/4,3T/4]$ the drive follows the analytic BBCD
protocol, and on $[3T/4,T]$ it is the time reverse of the first leg. We use five value, slope and curvature
constraints at $t=T/4$ which we eliminate linearly, so they
hold exactly for every coefficient vector $\bm{B}$:
\begin{align}\label{eq:hardcon}
  &g_i(T/4)=0,\quad \Delta=R_z(T/4),
  \nonumber\\
  &\dot g_i(T/4)=0,\quad \ddot g_i(T/4)=0 \quad \forall i,
\end{align}
The state is propagated on a uniform mesh of $N_t$ midpoint steps between $0$ and $T/4$ with time step
$\Delta t = T/(4N_t)$ and $t_j = j\Delta t$. The trotterized unitary evolution operator equals an ordered product of Rodrigues rotations, which are norm preserving. The discretized Eqs.~\eqref{veceom} and~\eqref{vecmag} are (using the midpoint rule, where $\bm{u}$ is evaluated at the midpoints $t_{j-1/2} = (j-1/2)\Delta t$, and $\hat{\bm{R}}$ at the mesh points)
\begin{align}\label{eq:dynamics}
  \hat{\bm{R}}(k,t_j) &= \mathrm{Rot}\!\big(2\Delta t\,\bm{u}(k,t_{j-1/2})\big)\, \hat{\bm{R}}(k,t_{j-1}), \qquad
 \nonumber\\
  R(k,t_j)   &= \bm{u}(k,t_{j-1/2}))\cdot\hat{\bm{R}}(k,t_{j-1}) ,
\end{align}
where 
\begin{equation}
\mathrm{Rot}(\bm{a})\bm{b} = \bm{a} + \frac{\sin|\bm{a}|}{|\bm{a}|}\bm{a}\times\bm{b} + \frac{1-\cos(|\bm{a}|)}{|\bm{a}|^2}\bm{a}\times(\bm{a}\times\bm{b})
\end{equation} is a rotation of $\bm{b}$ by an angle $|\bm{a}|$ about the axis $\bm{a}/|\bm{a}|$. We take $\hat{\bm{R}}(k,0) = \hat{\bm{R}}_{\rm RM}(k,0)$ and $\hat{\bm{R}}(k,t)$ is the unit Bloch vector of the filled band and
$R(k,t)=|\bm{R}(k,t)|$ is the instantaneous band energy (half the spectral gap). We use a mesh of $201$ $k$ points within $(-\pi/2a,\pi/2a)$. Our cost function minimized over $\bm{B}$ via Matlab's \texttt{particleswarm} and then
\texttt{fminunc} is
\begin{widetext}
\begin{align}\label{eq:cost}
  \mathcal{C}(\bm{B})
  &= \underbrace{\lambda_{\rm align}\sum_{k}
     \big\|\hat{\bm{R}}(k,t_{opt})-(0,0,-1)\big\|^{2}}_{\text{$T/4$ alignment}}
   + \underbrace{\left.\lambda_{\rm ks}\,\Delta k \sum_{k}
     \Big(\big(\delta_k^{2} \hat{R}_{x}\big)^{2}
        + \big(\delta_k^{2} \hat{R}_{y}\big)^{2}\Big)\right|_{t_{opt}}}_{k\text{-smoothness}}
     \nonumber\\&\quad+\underbrace{\lambda_{\rm ts}
     \sum_{i}^{5}\sum_{n=1}^{N_c} n^{4}
     \Big[(B^{\rm s}_{i,n})^{2}+(B^{\rm c}_{i,n})^{2}\Big]}_{\text{$t$-smoothness}}
   \nonumber\\
  &\quad + \underbrace{\frac{1}{N_t}\sum_{j=1}^{N_t}\sum_{k}\Big[
    \lambda_{\rm nv}\,\max\!\big(0,\,R_{\min}-s_{jk}\big)^{2}
  + \lambda_{\rm bd}\,\max\!\big(0,\,s_{jk}-R_{\max}\big)^{2}\Big]}_{\text{gap-bounding  terms}}\\[2pt]
\end{align}
\end{widetext}
where $s_{jk}\equiv\bm{u}(k,t_j)\cdot\hat{\bm R}(k,t_j)$, $t_{opt} = T/4$ and
$\delta_k^{2}f_k = (f_{k+1}-2f_k+f_{k-1})/\Delta k^{2}$. The second term is a penalty for having large contributions from higher $\varphi$-harmonics, and the final term bounds the gap of $\bm{R}$.

Figures~\ref{fig:ControlWeight} and~\ref{Q_pump} were produced using parameter values $a=1$, $J_0 = 1.1$, $\delta_0 = 0.9$, $\Delta_0 = 1$ and $\omega = 10$. For the bucket brigade leg of the protocol $t\in [\, T/4,3T)$, we choose $\theta(t) = 2\omega(t-T/4)-(1/2)\sin(4\omega(t-T/4))$ and $e^{\lambda(t)} = R^{opt}(T/4) + gb\sin^4(2\omega(t-T/4))$ where $R^{opt}$ is the value of $R(k,t)$ produced by the optimizer, at the end of the optimization and $gb = 3$. These both satisfy the boundary conditions (from $T/4$ to $3T/4$) for $\theta(t)$ and $\lambda(t)$ respectively.

\section{BBCD Robustness to Unknown Noise and Disorder}\label{appendix:robustness_unknown} \label{appendix:robustness_unknown_Htot} For the counterdiabatic bucket brigade protocol, in real space, the hopping coefficients and onsite potential can be found from Eqs.~\eqref{Htot} and~\eqref{eq:uCDk} via the inverse discrete Fourier transform, or directly from the real-space Hamiltonian~\eqref{eq:RMH-space} and are given by
\begin{align}\label{eq:Control_freak}
\begin{split}
    J_1+J^{1}_{CD} &= \left(e^{\lambda(t)}\sin{\theta(t)}-i\hbar\frac{\dot{\theta}(t)}{2}\right)\Theta\left(\frac{T}{2}-t\right)
    \\
    J_2+J^{2}_{CD} &= \left(e^{\lambda(t)}\sin{\theta(t)}+i\hbar\frac{\dot{\theta}(t)}{2}\right)\Theta\left(t-\frac{T}{2}\right)
    \\
    \Delta(t) &= e^{\lambda(t)}\cos{\theta(t)}
\end{split}
\end{align}
with the following time-boundary conditions on $\theta(t)$ and $\lambda(t)$: $\theta(t=0)=0$, $\theta(T/2)=\pi$ and $\theta(t=T)=2\pi$ and 
$\dot{\theta}(t=0) = \dot{\theta}(T/2)=\dot{\theta}(T) = 0$, as well as $\lambda(T) = \lambda(0)$. We impose spatially periodic boundary conditions:
\begin{equation}
    \begin{split}
        \ket{N+1,B} &= \ket{1,A}
    \end{split}
\end{equation}
Notice this gives rise to a dimerized system, and thus block diagonal total Hamiltonian $H$ at each instant in time. We now add spatial disorder and noise to $H(t)$:
\begin{widetext}
\begin{equation}\label{eq:timeDisNoise}
    \begin{split}
    H_{dis}(t)&=
        -\sum_{j=1}^{N}\left(\Delta^{z}_j(t)\ket{j}\bra{j}\otimes\sigma_z+\Delta^{0}_j(t)\ket{j}\bra{j}\otimes\sigma_0\right)-\sum_{m=1}^{N}\left( \delta J_1(t)\ket{m}\bra{m}\otimes 	\frac{\sigma_x-i\sigma_y}{2}+h.c\right) 
    \\
    &- \sum_{m=1}^{N-1}\left(\delta J_2(t)\ket{m+1}\bra{m}\otimes 	\frac{\sigma_x+i\sigma_y}{2} + h.c.\right)
    \end{split}
\end{equation}
\end{widetext}
where we further decompose these onsite terms into time-independent and dependent terms which correspond to the disorder and noise respectively, with
\begin{equation}\label{noiseDisorderDef}
    \Delta_j^i(t)= \eta \left(d^{i}\bar{\Delta}_j^i + n^{i}\Delta^i_{j,noise}(t)\right)
\end{equation}
where $\bar{\Delta}_j^i = \frac{1}{T}\int^Tdt \Delta_j^i(t)$ and thus $\bar\Delta^i_{j,noise}(t) = 0$ and $\eta$ is an overall strength of disorder/noise scale. The hopping noise is naturally proportional to the nominal strength of the hopping and thus for bucket brigade, $\delta J_2(t) = 0$ during the first half of the drive, and $\delta J_1(t) = 0$ during the second half of the drive and so the system remains dimerized for all $t$, so we take
\begin{equation}
\begin{split}
    \delta J_1(t)&=\eta \left(n^{x}\Delta^{x}_{j,noise}(t)\Theta(T/2-t)\right)
    \\
    \delta J_2(t)&=\eta \left(n^{y}\Delta^{y}_{j,noise}(t)\Theta(t-T/2)\right).
\end{split}
\end{equation} 
The parameters $d^{i}$ and $n^{i}$ tune the strength of the disorder and noise respectively. We assume spatially uncorrelated disorder, where the $\bar\Delta_j^i$ are chosen from the Gaussian distribution $N(0,1)$, with mean zero, and standard deviation one. We model the noise by convolving the square root of a Lorentzian (a spectral filter) with random variables drawn from $N(0,1)$. By construction, the resulting noise is time-periodic, and has time constant $t_c$.

Since our system involves non-interacting fermions, we can compute the single particle dynamics of the $N$ fermions, and then sum the results when computing the current. The system is initialized in its (many-body) ground state. It thus suffices to consider dynamics of each dimer subsystem initialized in its ground state: for the $j$th dimer, $\ket{\psi_j(0)} = \ket{j,A}$.

The bucket brigade protocol provides a further major simplification in that all of the dimers are decoupled. Thus, we can model the system's dynamics using the ``brick layer" method commonly used in quantum circuits. The unitary for each step of the bucket brigade protocol is given by
\begin{equation}
\begin{split}
    \hat{U}^i_j(t_f;t_0) &= \ket{j}\bra{j}\otimes\mathscr{T}\exp\left(-i \int^{t_f}_{t_0} dt H^i_j(t)\right)
    \end{split}
\end{equation}
where $j\in\{1,\dots N\}$ and $i\in\{1,2\}$. The $2\times2$ block matrices $H^{i}_j$ are given by
\begin{equation}\label{eq:HdisnoiseBBCD1}
    H^1_j(t) =
    \begin{pmatrix}
  -\Delta+\Delta^0_j-\Delta_j^z & -(J_1+J^{1}_{CD})^{*}-\delta J_1 \\
  -J_1-J^{1}_{CD}-\delta J_1 & \Delta+\Delta^0_j+\Delta_j^z
\end{pmatrix}
\end{equation}
which generates the dynamics for $0\leq t \leq T/2$ and
\begin{equation}\label{eq:HdisnoiseBBCD2}
    H^2_j(t) =
    \begin{pmatrix}
  \Delta+\Delta^0_j+\Delta_j^z & -(J_2+J^{2}_{CD})^{*}-\delta J_2 \\
  -J_2-J^{2}_{CD}-\delta J_2 & -\Delta+\Delta^0_{j+1}-\Delta_{j+1}^z
\end{pmatrix}
\end{equation}
which generates the dynamics for $T/2< t \leq T$. The first leg of the protocol takes a particle initially localized at a A sublattice to the B sublattice point in the same unit cell. We implement the second leg that takes a particle from the B sublattice in one unit cell to the A sublattice of the next unit cell by shifting the B and A sublattices into one another.
\begin{figure*}[t]
    \centering
        \includegraphics[width=1.0\textwidth]{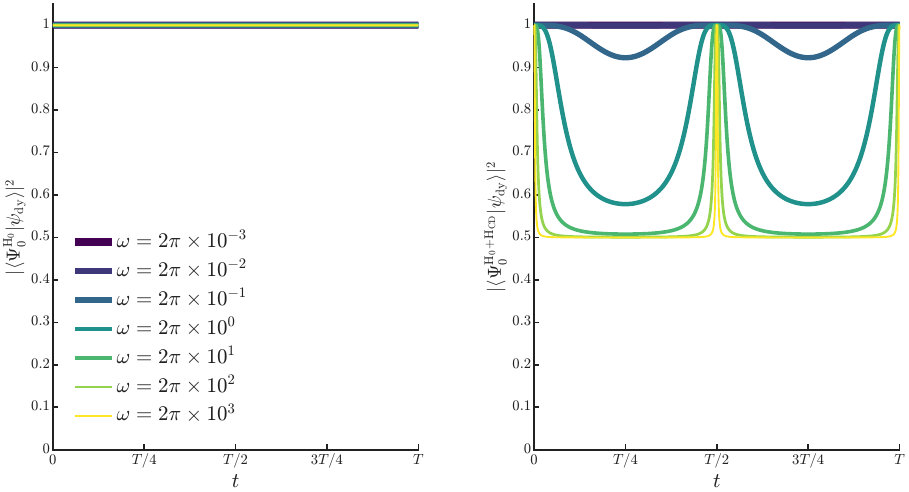}
    \caption{\textbf{Overlaps of dynamic state and the ground state manifold of $H_0$ and $H_{tot}$}. We manifestly see that for speeds larger than adiabatic with respect to $H_0$, CD driving carries the dynamic state only along the ground state manifold of $H_0$ rather than the ground state manifold of $H_{tot}=H_0+H_{CD}$}
    \label{fig:H0vsCDoverlaps}
\end{figure*}
Thus, we can use the brick-layer method to produce the dynamics for each dimer. The total evolution operator is thus:
\begin{equation}
    \hat{U}(t\leq T/2;0) = \prod_j \hat{U}^{1}_{j}(t;0)
\end{equation}
\begin{equation}
\begin{split}
    \hat{U}(t\geq T/2;\,T/2) 
    &= \hat{cs}^{-1}\left(\prod_{j'}\hat{U}_{j'}^{2}(t;T/2)\right)\hat{cs} \\
    &\quad\cdot\left(\prod_j \hat{U}^{1}_{j}(T/2;0)\right)
\end{split}
\end{equation}
where 
\begin{equation}
\begin{split}
    \hat{cs} = \sum_{n}\biggl(\ket{n A}\bra{nB}+\ket{(n \text{ Mod } N+1 )B}\bra{nA}\biggr)
    \end{split}
\end{equation} is the circular shift permutation operator which the lattice by one site, which maps the A sublattice of $n$th unit cell into the B sublattice of the same cell and the B sublattice of $n$th cell into the A sublattice of the $n+1$ cell. By construction, $\hat{cs}$ thus implements periodic boundary conditions on the index $j$ in the Hamiltonian Eq.~\eqref{eq:HdisnoiseBBCD1} and ~\eqref{eq:HdisnoiseBBCD2}. We can furthermore simplify the propagator by choosing $\theta(t)$ such that $J_1(t)$ over the interval $T/2 \leq t \leq T$ coincides with $J_2(t)$ over $0 \leq t \leq T/2$. 
Since our system is a set of dimers, the integrated current through all bonds is given by the difference of the charge on the sites straddling each bond, i.e. the polarization, or equivalently, the dipole moment of each dimer as in Eq.s~\eqref{eq:dimerContEq1} and~\eqref{eq:dimerContEq2} of the main text.

We plot the distribution of values $C$ for a range of drive speeds $
\omega$, disorder and noise strengths, in Fig.~\ref{fig:RobustTrip} of the main text. Figure was produced with the 100 realizations of the noise/disorder using the following parameter and function choices: $N =100$ unit cells, $\theta(t) = \omega t -(1/2)\sin(2\omega t)$, where $\omega = 2\pi/T$, and $\lambda(t) = 0$, and so gap of $H_0(t)$ is equal to $1$, which determines the adiabatic limit.  We also take the time constant $t_{c}= 1$ and arbitrarily choose the noise and disorder parameters to be $n^{0} = 0.8, n^{z} = 1.0, n^{x} = 1.3, n^{y} = 1.5, d^{0} = 0.9, d^{z} = 1.2$. For the probability distribution KDE, we use a bandwidth floor of $h_{min}= 0.008$; so that the peak heights of the delta functions are $1/(h_{min}\sqrt{2\pi}) \approx 50$. We note that the overlap plots show that for long cycles (slow driving) and large noise the overlap is near $0.5$: the system's state settles into a randomization of dimer occupancy where a nearly equal number of dimers are in their ground and excited states, rather than all remaining in their ground states.

\section{BBCD Robustness to Known, Static Disorder}\label{appendix:robustness_known_dis} The numerics in the main text again use the brick layer method to evolve the system and compute the state-transfer fidelity (overlap) and pumped charge distribution plots of Fig.~\ref{fig:RobustDisCD} for $100$ realizations of the disorder in an $N = 100$ chain. Plots produced using the parameter values $d^z = 0.99$ and $d^0 = 0.96$. In the main text, we present an analytic method which directly harnesses the physical picture of the BBCD protocol; here, we also present an algebraic proof as well. We introduce weak, onsite disorder to the bucket-brigade $H_0(t)$ as described in the main text, Eq~\eqref{eq:DisHam}. This breaks discrete translation invariance, so $k$ is no longer a good quantum number. However, since the bucket-brigade protocol leads to a dimerized, or monomerized system at all times, the two legs of the protocol are given by the following set of decoupled $2\times2$ block matrices $H^{i}_j$:
\begin{widetext}
\begin{equation}\label{eq:HdisBBCD1}
    H^1_j(t) =
    \begin{pmatrix}
  -\Delta+\eta\left(d^0\Delta^0_j- d^z\Delta_j^z\right) & -J_1 \\
  -J_1 & \Delta+\eta\left(d^0\Delta^0_j+d^z\Delta_j^z\right)
\end{pmatrix}
\end{equation}
which generates the dynamics for $0\leq t \leq T/2$ and
\begin{equation}\label{eq:HdisBBCD2}
    H^2_j(t) =
    \begin{pmatrix}
  \Delta+\eta\left(d^0\Delta^0_j+d^z\Delta_j^z\right) & -J_2 \\
  -J_2 & -\Delta+\eta\left(d^0\Delta^0_{j+1}-d^z\Delta_{j+1}^z\right)
\end{pmatrix}
\end{equation}
which generates the dynamics for $T/2< t \leq T$.
\end{widetext}
We can write each of these $2\times2$ matrices as $A^{i}\sigma_0+\bm{R}^{i}_{j}(t)\cdot\bm{\sigma}$. In particular, 
\begin{equation}\label{eq:RBBdis}
\begin{split}
    \bm{R}^{1}_{j}(t)&=(-J_1(t),0,-\left(\Delta(t)+\eta d^{z}\Delta^{z}_j\right))
    \\
    \bm{R}^{2}_{j}(t)&=(-J_2(t),0,\Delta(t)+D_{j,j+1}).
    \end{split}
\end{equation}
where $D_{j,j+1} = \frac{\eta}{2}\left(d^0\left(\Delta^{0}_j-\Delta^{0}_{j+1}\right)+d^z\left(\Delta^{z}_{j}+\Delta^{z}_{j+1}\right)\right)$. We define $\mathcal{D}^1_j = d^z\Delta^z_{j}$ and $\mathcal{D}^2_j = D_{j,j+1}$. We can immediately write down the CD operator for each dimer $j$ and each leg of the protocol $i = 1,2$ again using the form of the CD operator for a $2\times 2$ Hamiltonian: $H^{i}_{CD,j}(t) = \frac{\hbar}{2|R_j^i|^2}\bm{R}^{i}_j\times\partial_t \bm{R}^{i}_j\cdot \bm{\sigma}$, as in Eq~\eqref{Htot}:
\begin{equation}
\begin{split}
    H^{i}_{CD,j}(t) &= s^{i}b^{i}_j(t)\sigma_y
    \\
    b^{i}_j(t) &= \frac{\hbar}{2}\,
   K\!\left(\theta,\dot{\theta},\lambda,\dot{\lambda};\mathcal{D}^{(i)}_{j}\right)
    \end{split}
\end{equation}
where $s^{1}=1,s^{2}=-1$ and
\begin{equation}
    K\!\left(\theta,\dot{\theta},\lambda,\dot{\lambda};\mathcal{D}^{(i)}_{j}\right)=\frac{\dot{\theta}+\varepsilon_{i,j} \left( \dot{\theta}\cos\theta+\dot{\lambda}\sin\theta\right)}
{\big(1+2\varepsilon_{i,j} \cos\theta+\varepsilon^{2}_{i,j}\big)},
\end{equation}
with $\varepsilon_{i,j}(t) \equiv\mathcal{D}^{(i)}_{j}/e^{\lambda(t)}$.
Moreover, we can also write down the ground state density matrix for each dimer and each part of the protocol: $\rho^{i}_j(t) = \frac{1}{2}\left(\mathbb{I} - \hat{\bm{R}}^{i}_j(t)\cdot\bm{\sigma}\right)$. These suffice to analytically compute the pumped charge using Eq.~\eqref{eq:javggen} in analogy with the clean, BBCD, $k$-space results. So starting from:
\begin{equation}
j_{\rm avg}(t)=-\frac{1}{2N}\frac{ie}{\hbar}
\sum_{x,x'}f\!\Big(\frac{x-x'}{a}\Big)\,H(x,x',t)\,\rho(x',x,t)
\end{equation}
we have $f(\pm1)=\pm1$ for
nearest neighbors, and using Hermiticity of $H$ and $\rho$ and dimerization of the BB drive we find:
\begin{equation}
j^{i}_{\rm avg}(t)=\frac{2e}{N\hbar}\sum_{j=1}^{N}
\mathrm{Im}\!\left[H^{i}_{j,21}\,\rho^{i}_{j,12}\right].
\end{equation}
Because both $\rho^{i}_{j,12}$ and $H^{i}_{j,21}$ are real
only the CD bond carries current in the instantaneous
ground state:
$\mathrm{Im}[H_{21}\rho_{12}]_{i,j}
=(s^{i}b^{i}_{j}(t))(1/2)\left(\sin\theta/\left(1+2\varepsilon_{i,j} \cos\theta+\varepsilon^{2}_{i,j}\right)^{1/2}\right)$.
We thus find,
\begin{equation}
j^{(i)}_{\rm avg}(t)= s_{i}\,\frac{e}{2N}\;\sum_j F\!\left(u,\varepsilon_{i,j}\right).
\end{equation}
where 
\begin{equation}
    F\!\left(u,\varepsilon_{i,j}\right) = \frac{-\dot{u}+(1-u^2)\dot{\varepsilon}_{i,j}+(1/2)\varepsilon_{i,j}\partial_t \left(u^2\right)}{\left(1+2u\varepsilon_{i,j}+\varepsilon_{i,j}^2\right)^{3/2}}
\end{equation}
where $u= \cos\theta(t)$ and so $\dot{u} = -\dot{\theta}\sin\theta$ as well as $\dot{\varepsilon}_{i,j} = -\dot{\lambda}\varepsilon_{i,j}$. We now show that $F\!\left(u,\varepsilon_{i,j}\right)$ is an exact differential (exact one-form on the $(u,\varepsilon)$ space). Define 
\begin{equation}
    G(u,\varepsilon_{i,j}) =\frac{\varepsilon_{i,j}+u}{\left(1+2u\varepsilon_{i,j}+\varepsilon_{i,j}^2\right)^{1/2}}
\end{equation}
then using $\frac{dG}{dt} = \partial_uG\,\dot{u}+\partial_{\varepsilon_{ij}}G\,\dot{\varepsilon}_{ij}$, we see that
\begin{equation}
  \frac{dG(u,\varepsilon_{i,j})}{dt} = -F\!\left(u,\varepsilon_{i,j}\right)
\end{equation}
Hence, the average pumped charge, $Q_{pump}(t) = \int dt (1/2)\sum_ij^{i}_{\rm avg}(t)$, is a $(u,\varepsilon_{i,j})$-path independent quantity. Moreover, during leg 1, $\sin\theta\geq 0$, and thus $u: 1 \to -1$ while over leg 2, $\sin\theta\leq 0$, and thus $u: -1 \to 1$. Hence, we find:
\begin{equation}
    \Biggr. G\!\left(u,\varepsilon_{1,j}\right)\Biggl|^{{T/2}}_{0} = -sgn(1-\varepsilon_{1,j}(T/2))-sgn(1+\varepsilon_{1,j}(0))
\end{equation}
and similarly 
\begin{equation}
    \Biggr. G\!\left(u,\varepsilon_{2,j}\right)\Biggl|^{{T}}_{T/2} = sgn(1+\varepsilon_{2,j}(T))+sgn(1-\varepsilon_{2,j}(T/2))
\end{equation}
 and so, the average pumped charge is
\begin{widetext}
\begin{equation}
\begin{split}
      Q_{pump}(t) &= \frac{e}{4N}\sum_{j=1}^{N}\Biggl(sgn\biggl(e^{\lambda(T/2)}-\eta d^z\Delta^z_j\biggr)+sgn\biggl(e^{\lambda(0)}+\eta d^z\Delta^z_j\biggr)
      \\&+sgn\biggl(e^{\lambda(T)}+D_{j,j+1}\biggr)+sgn\biggl(e^{\lambda(T/2)}-D_{j,j+1}\biggr)\Biggr)
\end{split}
\end{equation}
\end{widetext}
 and thus we see the average pumped charge remains exactly quantized, away from the adiabatic limit for weak disorder. Interestingly, only the gap at end points of each leg of the protocol determine quantization. In the main text, we derived the same threshold conditions on the disorder to preserve pumped charge quantization, Eq.~\eqref{eq:sync_cond}, by considering the continuity equation of each dimer, and noting that quantization of pumped charge is guaranteed so long as the disorder does not spoil the synchronization of dimers' initial and final states of the protocol legs.

\section{Counterdiabatic Driving and adiabaticity of the total Hamiltonian}\label{appendix:H0vsCD}
A common perspective for how CD driving reproduces adiabatic dynamics in finite time is that the dynamics of the total system are adiabatic with respect to $H = H_0 + H_{CD}$, while being fast compared to $H_0$, since the CD term contributes significantly to the gap of $H$~\cite{abah2019energetic,duncan2025taming}. In other words, imagine the time scale of driving is given by $T \ll \hbar/ \Delta E_0$, where $\Delta E_0$ is the gap of $H_0$. Hence the system is driven non-adiabatically with respect to $H_0$. The common view is that, in contrast, under CD driving, this same $T \gg \hbar / \Delta E$, where $\Delta E$ is the gap of $H = H_0 + H_{CD}$, so the system evolves adiabatically with respect to the total $H$. However this argument cannot be the explanation for how CD works, because adiabaticity for $H$ would mean the system follows the ground state manifold of $H$. By construction, we know a CD-driven system follows the ground state manifold of $H_0$. But the ground state manifold of $H_0$ is generically not the ground state manifold of $H$, unless we are driving extremely slowly.
Numerically, we can straightforwardly demonstrate the above argument with a two-dimer, clean Thouless pump using our BBCD protocol.  Suppose we drive at a speed $\omega = \frac{2\pi}{T}$ with cycle time $T \ll \frac{\hbar}{ 2\mathcal{E}e^{\lambda(t)}}$, while $T \gg \frac{\hbar} {2\sqrt{\mathcal{E}e^{2\lambda(t)}+\hbar^2\dot{\theta}^2(t)/4}}$, where $\dot\theta \propto\omega$ and $\mathcal{E}$ has units of energy. We take $\frac{\hbar}{\mathcal{E}} = 1$. We see from the right panel of Fig.~\ref{fig:H0vsCDoverlaps} that the overlap of the dynamic state and the ground state manifold of $H_0 + H_{CD}$ is not $1$, for speeds away from the adiabatic limit (which for our parameter choices is roughly $T \sim 10^2 \iff \omega \sim 10^{-2}$). Only in the adiabatic limit does $H \approx H_0$ and then the ground state manifolds coincide.

\end{document}